\documentclass[pra,twocolumn,
superscriptaddress,
longbibliography,
showpacs,
aps
]{revtex4-2}
\usepackage{graphicx}
\usepackage{amsmath}
\usepackage{amsfonts}
\usepackage{upgreek}
\usepackage{amssymb}
\usepackage[colorlinks=true,linkcolor=blue,urlcolor=blue,citecolor=blue]{hyperref}
\usepackage{color}
\usepackage{lipsum} 
\newcommand{\ket}[1]{|{#1}\rangle}
\newcommand{\bra}[1]{\langle{#1}|}

\newcommand{\beq}{\begin{equation}}
\newcommand{\eeq}{\end{equation}}

\usepackage{dsfont}
\usepackage{placeins} 
\usepackage{graphicx}
\usepackage{wrapfig}
\usepackage{siunitx}
\usepackage{dcolumn}
\usepackage{bm}
\usepackage{orcidlink}

\begin{document}

\title{Continuous cavity-QED with an atomic beam}

\author{Francesca Famà\,\orcidlink{0009-0009-3162-8806}}
\affiliation{Van der Waals-Zeeman Institute, Institute of Physics, University of Amsterdam, Science Park 904,
1098XH Amsterdam, The Netherlands}

\author{Sheng Zhou} 
\affiliation{Van der Waals-Zeeman Institute, Institute of Physics, University of Amsterdam, Science Park 904,
1098XH Amsterdam, The Netherlands}

\author{Benedikt Heizenreder} 
\affiliation{Van der Waals-Zeeman Institute, Institute of Physics, University of Amsterdam, Science Park 904,
1098XH Amsterdam, The Netherlands}

\author{Mikkel Tang} 
\affiliation{Van der Waals-Zeeman Institute, Institute of Physics, University of Amsterdam, Science Park 904,
1098XH Amsterdam, The Netherlands}
\affiliation{Niels Bohr Institute, University of Copenhagen, Blegdamsvej 17, 2100 Copenhagen, Denmark}

\author{Shayne Bennetts} 
\affiliation{Van der Waals-Zeeman Institute, Institute of Physics, University of Amsterdam, Science Park 904,
1098XH Amsterdam, The Netherlands}

\author{Simon B. J\"{a}ger\,\orcidlink{0000-0002-2585-5246}}
\affiliation{Physics Department and Research Center OPTIMAS, University of Kaiserslautern-Landau, 67663 Kaiserslautern, Germany}

\author{Stefan A. Sch\"{a}ffer\,\orcidlink{0000-0002-5296-0332}}
\affiliation{Van der Waals-Zeeman Institute, Institute of Physics, University of Amsterdam, Science Park 904,
1098XH Amsterdam, The Netherlands}
\affiliation{Niels Bohr Institute, University of Copenhagen, Blegdamsvej 17, 2100 Copenhagen, Denmark}

\author{Florian Schreck\,\orcidlink{0000-0001-8225-8803}}
\email[]{continuousCollectiveGain@strontiumBEC.com}
\affiliation{Van der Waals-Zeeman Institute, Institute of Physics, University of Amsterdam, Science Park 904,
1098XH Amsterdam, The Netherlands}
\affiliation{QuSoft, Science Park 123, 1098XG Amsterdam, The Netherlands}

\date{\today}

\begin{abstract}
Atoms coupled to cavities provide an exciting playground for the study of fundamental interactions of atoms mediated through a common channel. 
Many of the applications of cavity-QED and cold-atom experiments more broadly, suffer from limitations caused by the transient nature of an atomic loading cycle. The development of continuous operation schemes is necessary to push these systems to the next level of performance. 
Here we present a machine designed to produce a continuous flux of collimated atoms that traverse an optical cavity. The atom-light interaction is enhanced by a fast-decaying cavity optimal for studying phenomena where atomic properties dominate. We demonstrate the transition to a collective strong coupling regime heralded by a normal-mode splitting. We observe a second phase with a binary normal-mode splitting born from an offset in the mean velocity of the atoms. Inverting the atomic ensemble in the collective strong coupling regime, we measure continuous optical gain. This work sets the stage for studying threshold conditions for continuous collective phenomena, such as continuous superradiant lasing.

\end{abstract}

\maketitle

\section{\label{sec:1}Introduction}
Developing techniques to realize the continuous operation of quantum systems based on cold atoms is a long-standing goal within the quantum community~\cite{Dieckmann1998, Lahaye2005, Chen:2022, Huntington2023, Okaba:2024,schafer:2024,Gyger2024}. This requires that atoms are continuously loaded into the region of interest, as their typical trapping times are usually limited by heating and decoherence processes~\cite{Norcia2024}. Consequently, one of the main challenges arises from the finite interaction time of the individual atoms and whether it fundamentally limits the coherence time of the quantum system~\cite{Acin:2018}. Overcoming this challenge will result in the realization of continuous cold atom platforms, which will find broad applications in devices including quantum simulators and computers~\cite{Xu:2024}, but also ultra-precise quantum sensors and clocks~\cite{Olson2019,Katori2021,Takeuchi2023}.

A paradigmatic example of such a device is a continuous beam of cold atoms that traverses and interacts with an optical cavity, which can mediate interaction between the atoms. For cavity decay rates much larger than the other rates involved, the cavity field will have a short memory of phase information. This case is often referred to as the 'bad-cavity' regime, in which the cavity acts as a dissipative entity, providing a preferred vacuum channel to which all atoms couple \cite{Carmichael88}. Especially interesting is the case of a narrow, semi-forbidden atomic transition couples to the optical cavity. Such a coupled system might be used to realize the next generation of superradiant lasers and active atomic clocks that operate in continuous-wave mode~\cite{Liu:2020, Tang:2022}. Devices working in the bad-cavity regime, can be extremely robust against environmental noise~\cite{Numata2004, Meiser:2009}. Furthermore, they can overcome the issue of discontinuous phase evolution emerging from cyclic operation, seen, e.g., as the Dick effect, which limits the frequency stability of state-of-the-art atomic clocks~\cite{Dick1987, Westergaard2010}. The phase-noise introduced through cycling significantly affects the time scaling by limiting the $1/\text{time}$ scaling of white phase noise to $1/\sqrt{\text{time}}$ at all operating time scales. On the fundamental side, such systems are formidable playgrounds for exploring many-body physics emerging from the interplay of continuous atom flow and long-range cavity-mediated atom-atom interactions~\cite{Kim2017}. Collective atom-cavity strong coupling has been observed in (quasi-)continuous systems for narrow transitions with ultracold atoms in a cavity \cite{Kristensen:2023, Norcia:2018:2,cline:2022}, and in thermal atoms transiting a cavity mode, with broad atomic transitions coupled to the same mode \cite{gea2008}. However, it remains an open question whether the cavity and atoms can form steady-state coherent dressed states for a semiforbidden atomic transition, having to overcome the finite lifetimes of cavity photons and atoms. For an atomic beam, this is particularly challenging as the lifetime of the atoms in the cavity mode is determined by their finite transit time rather than their natural, narrow linewidth. The demonstration that coherent collective atom-cavity interactions can overcome these processes is an important step towards establishing such devices as future quantum technologies \cite{zhang2023}.

In this paper, we demonstrate the formation of dressed states of a thermal atomic beam that traverses an optical cavity, see Fig.~\ref{Fig:1}(c). In particular, we report strong collective light-matter coupling of the narrow $7.5$~kHz $^1S_0$$\leftrightarrow$ $^3P_1$ transition of $^{88}$Sr to a bad cavity. The transit times of the atoms is the order of hundreds of nanoseconds, while the cavity field lifetime is of the order of few nanoseconds. In this regime, the observation of normal-mode splitting, that heralds the formation of dressed states, requires high atomic fluxes and fast, effective collimation and cooling stages, both of which we have realized in our experiment. We provide an extensive overview of the experiment and theory used to analyze the presented atomic-beam cavity setup. In particular, we observe that small atomic beam deflections result in modified atom-light states in the cavity mode, which materializes as a binary normal-mode splitting. We also discuss the requirements for realizing a continuous-wave superradiant laser with this setup, potentially leading to an atom-based laser with an inherent Hz-level linewidth \cite{Tang:2022}. In this context, we demonstrate gain in our system and discuss future steps required to enter the lasing regime with our setup. Our work is an important step towards realizing quantum technologies based on cold atomic beams interacting with an optical cavity.

\section{\label{sec:2}Experimental apparatus}
Our system consists of a continuous thermal beam of $^{88}$Sr atoms that traverses an optical resonator. During transit, the atomic transition between the ground state $\ket{g} = \ket{{}^1S_0}$ and the excited state $\ket{e} = \ket{{}^3P_1, m_j=0}$ couples to the resonant TEM$_{00}$-mode of the optical cavity with linewidth $\kappa=2\pi\times\qty{53.9(2)}{MHz}$. The atomic transition has a wavelength of $\lambda =$~\qty{689}{nm} and a natural linewidth $\gamma=2\pi\times$~\qty{7.48(1)}{kHz}~\cite{nicholson:2015}, placing our experiment deep in the bad-cavity regime, $\gamma\ll\kappa$. The remaining cavity parameters are summarized in Tab.~\ref{Table:1}, in particular, the single-atom cavity coupling $g_0=2\pi\times\qty{22.4}{kHz}$ and cooperativity $C=1.2\times10^{-3}$. The purpose of our experiment is to achieve strong collective atom-cavity interactions and for this we need to meet a threshold of mean intra-cavity atom number $N$ and reduce several severe broadening mechanisms inherent to atomic beam setups. 
Inhomogeneous broadening is mainly due to the velocity distribution of the atoms along the cavity axis. This results in Doppler broadening described by $\delta_D=k\Delta v$ and is reduced by cooling the atoms before entering the cavity. In the experiment, we typically access Doppler broadenings of $\delta_D\sim2\pi\times\qty{5}{MHz}$. Homogeneous broadening of the atoms is collected in the atomic dephasing rate $1/T_2$ that have multiple origins. The main contributions to the latter are the free-space spontaneous photon emission with rate $\gamma$, the stray light described by the rate $\gamma_B$ and the finite transit times $\tau$ that lead to transit-time broadening. The stray light is caused by blue laser cooling light, below the cavity, scattered by atoms into the cavity region. Both the rates $\gamma$ and $\gamma_B$ are on the order of $\qty{10}{kHz}$ while the typical transit time broadening is of the order of\,MHz, making it the main source of homogeneous broadening in our setup. We include in our description the Maxwell-Boltzmann distribution of atomic velocities along the beam propagation axis, which results in varying transit-time broadenings. 

\begin{figure}[h!]
\includegraphics[width=1\linewidth]{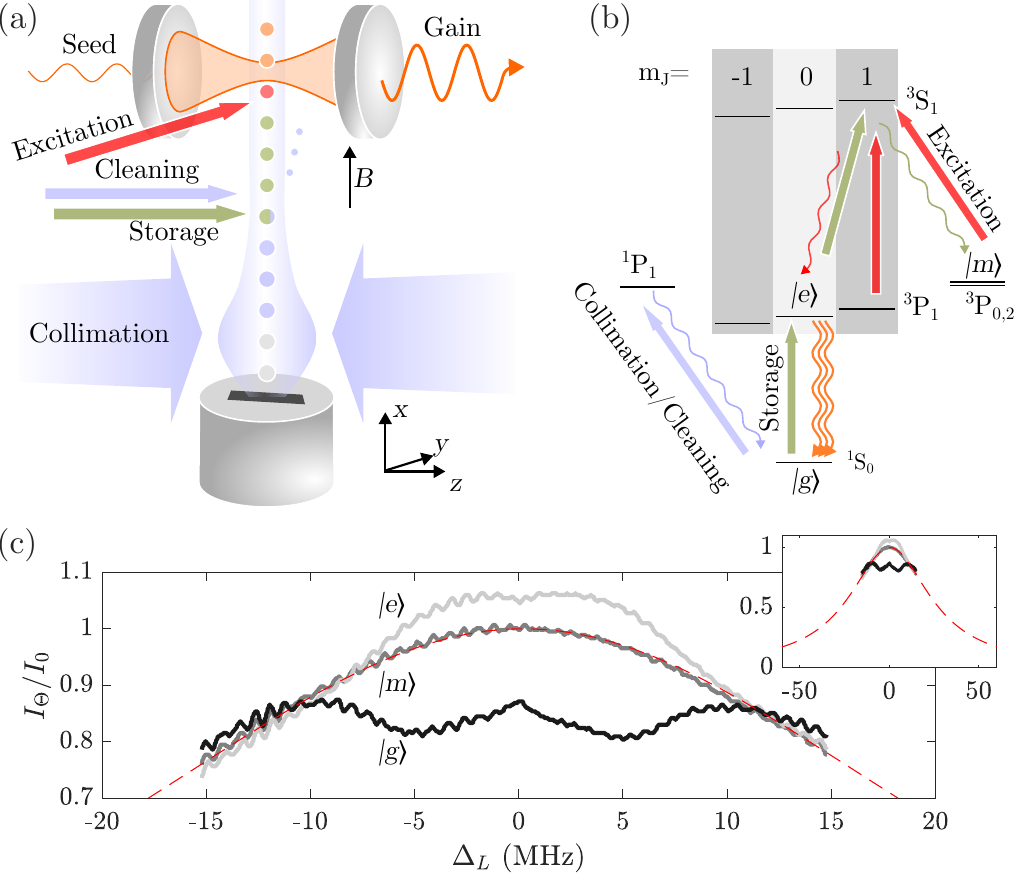}
\caption{\label{Fig:1} (a) Schematic of the apparatus. An oven emits an atomic beam of $^{88}$Sr atoms through a nozzle. The beam is then collimated by an optical molasses. Next, the atoms are transferred from the ground state $\ket{g} = {}^1S_0$ to the upper lasing state $\ket{e} = {}^3P_{1, m_j = 0}$ through a 3-step process: 1) storage, by optical pumping to the metastable states $\ket{m} = {}^3P_{0, 2}$; 2) cleaning, by accelerating the remaining ${}^1S_0$ atoms to velocities along the cavity axis at which they are irrelevant; and 3) excitation, by optically pumping atoms from $\ket{m}$ to $\ket{e}$. Finally, the prepared sample traverses the TEM$_{00}$-mode of a bad-cavity optical resonator. (b) Relevant level structure of the Sr atoms. We specify wavelenghts and polarizations used in the collimation and 3-step sequence. For clarity, some optical pumping transitions are not shown. (c) Examples of transmission measurements as a function of probe detuning with respect to the $\ket{g} \rightarrow \ket{e}$ transition. For atoms in ${}^1S_0$ (dark, $\Theta=g$), ${}^3P_{2,0}$ to mimic an empty cavity (medium, $\Theta=m$), and ${}^3P_{1, m_j = 0}$ (light, $\Theta=e$). A Lorentzian distribution representing the ideal empty cavity is shown with a dashed red line and is used for normalization  (see Sec.~\ref{sssec:2.1E})}
\end{figure}

Before the atoms enter the cavity, we use optical electronic transitions to spatially collimate, cool, and state prepare the atoms; see Fig.~\ref{Fig:1}(b). This serves to reduce Doppler broadening and increase the mean number of intra-cavity atoms $N$. A magnetic field ${\bf B} = B{\bf e}_x$, with $B = 1.5$~G, sets the quantization axis along the atomic beam propagation $x$-axis and lifts the $m_j$ degeneracy of the ${}^3P_1$ state.
The preparation takes place through a three-stage process; see Fig.~\ref{Fig:1}(a), designed to realize an ensemble that satisfies the requirements for generating collective states of ground~\cite{Wickenbrock2011, Thompson1992} or excited-state atoms~\cite{Liu:2020}. In particular, with our preparation scheme, we continuously prepare up to \num{1e6} atoms in $\ket{g}$ and up to \num{6e5} atoms in $\ket{e}$ that couple to the cavity mode volume at any given time. We control the final atomic state by moving through the metastable, stretched, fine-structure triplet states $\ket{m} = \ket{{}^3P_{0,2}}$. They do not couple to the cavity mode and we can therefore use them to mimic the empty cavity. We thus realize the strong collective atom-cavity coupling regime on the $\ket{g} - \ket{e}$ transition. 

\begin{table}[b]
\centering
\begin{tabular}{|p{5cm} || c| l|} 
 \hline
  Cavity Linewidth & $\kappa/2\pi$ & $53.9(2)$~MHz\\
 \hline
  Free Spectral Range & FSR$/2\pi$ & $5.479(3)$~GHz\\
 \hline
 Finesse & $F$ & $101.6(8)$ \\ 
 \hline
 Mode Waist Radius & $w$ & $86.9(2)~\upmu$m\\ 
 \hline
 Single Atom Cooperativity & $C$ & $1.24(1)\times10^{-3}$ \\
  \hline
 Cavity Length & $L$ & $27.38(1)$~mm\\
 \hline
 Single Atom Coupling Rate & $g_0/2\pi$ & $22.4(3)$~kHz\\ 
 \hline
\end{tabular}
\caption{Key cavity parameters. The cavity linewidth $\kappa$ and the free spectral range are measured. The remaining quantities are derived from these values and the specified mirror radius of curvature of $100.0(1)$~mm.} \label{Table:1} 
\end{table}

In the following paragraphs, we elaborate on the experimental details that are required to describe our system.

\subsubsection{\label{sssec:2.1A} Velocity profile along the beam axis}
\begin{figure}
\centering
\includegraphics[width=1\linewidth]{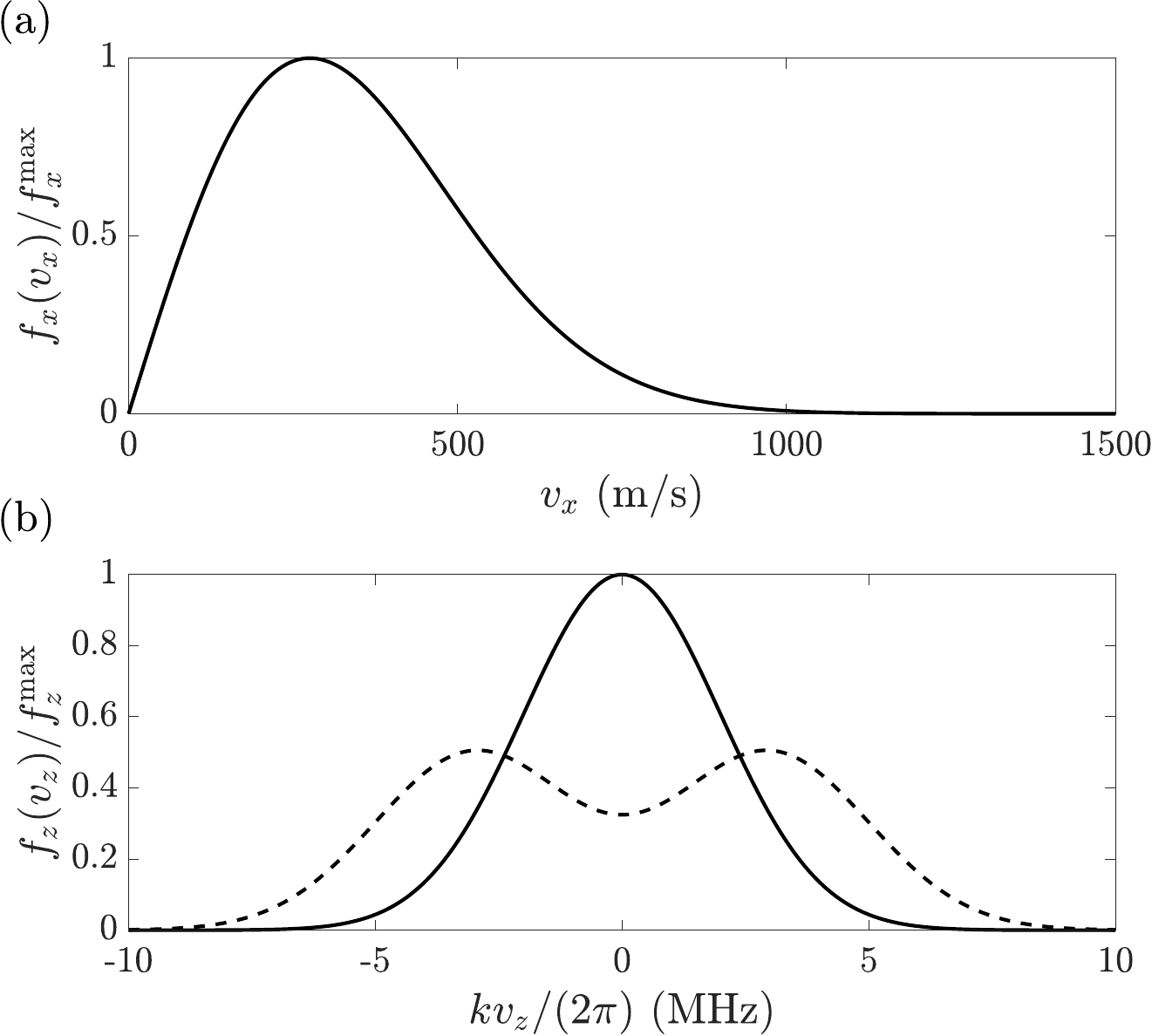}
\caption{\label{Fig:2} (a) Velocity distribution $f_x$ in Eq.~\eqref{eq:vxdistr} normalized to its maximum, plotted for $T=\qty{803}{K}$. (b) Velocity distribution $f_z$ in Eq.~\eqref{eq:vzdistr} normalized by its maximum as function of $kv_z$ for  $\delta_D=2\pi\times\qty{2}{MHz}$. $\delta_0=kv_{z,0}=0$ (solid) and $\delta_0=2\pi\times\qty{3}{MHz}$ (dashed).}
\end{figure}

Atoms are emitted upward directly from an oven with a micro-capillary tube array (305~mm inner diameter, 414~mm outer diameter, 8~mm length) arranged in a rectangular opening (dimensions $l_z = 11$mm and $l_y = 2$mm). This nozzle is elongated to take advantage of the cylindrical mode volume of the cavity and increases the overlap volume, see Fig.~\ref{Fig:1}(a). No manipulation is performed along the vertical axis ($x$-axis).
To estimate the transit time as a function of oven temperature $T$, we model the atomic beam well below the Knudsen diffusion regime with a velocity distribution along the $x$-axis of
\begin{equation}
    f_{x}(v_x) = \beta_x m v_x e^{-\beta_x\frac{m v_x^2}{2}},\label{eq:vxdistr}
\end{equation}
where $\beta_x= 1/k_B T$. The mean velocity is given by
\begin{align}
    \bar{v}_x = \int_{0}^{\infty} dv_x\,v_xf_x(v_x)=\sqrt{\frac{\pi}{2m\beta_x}}, \label{eq:vxmean}
\end{align}
resulting in a mean transit time through the cavity waist of
\begin{align}
\tau=\frac{2w}{v_x} = \left(\frac{8w^2m\beta_x}{\pi}\right)^{-1/2},\label{eq:tau}
\end{align}
where $w$ is the waist of the optical cavity [see Tab.~\ref{Table:1}].
In Fig.~\ref{Fig:2}(a) we show an example of the velocity distribution in Eq.~\eqref{eq:vxdistr} for $T=\qty{803}{K}$, where the mean velocity is $\bar{v}_x\approx\qty{345}{m/s}$ and the corresponding mean transit time is $\tau\approx~\qty{0.5}{\upmu s}$.  
We confirm the general features of the modeled velocity distribution given by Eq.~\eqref{eq:vxmean} by performing direct time-of-flight measurements on the system, see App.~\ref{App:B}.

\subsubsection{\label{sssec:2.1B}Velocity selective storage} 

In the transverse plane, our aim is to reduce the Doppler width along the cavity ($z$-axis) and reduce the spread of the atomic beam along the $y$-axis. 
We use optical molasses on the broad (30-MHz linewidth) blue ${}^1S_0\rightarrow {}^1P_1$ transition, along the $y$-axis and $z$-axis over a distance of $\qty{5}{cm}$. The detuning of the lasers is -14\,MHz along $y$ and -16\,MHz along $z$. Intensities of $1.8~I_{\mathrm{sat}}$ along the $y$-axis and $5~I_{\mathrm{sat}}$ along the $z$-axis are used, where the saturation intensity is $I_{\mathrm{sat}} =41$~mW/cm$^2$. Even at these high intensities, the set of two $z$-axis molasses beams experience $55$ and $70$\% ($70$ and $90$\%) absorption respectively at an oven temperature of \qty{803}{K} (\qty{833}{K}), due to the optical density of the atomic beam. This leads to a $z$-dependent intensity imbalance between the counter-propagating molasses beams, and therefore position-dependent forces, resulting in a broader Doppler distribution than expected from a balanced system. Along the $y$-axis the atomic beam is thinner and the beam intensity imbalance is negligible. 
A residual angle between the atomic beam and the $x$-axis in the cavity plane can originate from a tilt of the oven nozzle and imperfect molasses beams.
Small angles can quickly manifest as noticeable velocity shifts in the cavity mode (e.g., a 0.5-degree shift corresponds to \qty{3.5(5)}{m/s}). 
The significant absorption of our beams at high temperatures makes correcting such angles with the molasses challenging. For this reason, we typically operate at maximum temperatures around \qty{803}{K}.
This atomic misalignment is modeled by introducing an offset in the velocity distribution along the cavity axis.

As photons can travel in two opposing directions when coupling to the cavity mode, the resulting effective velocity profile coupling to the intra-cavity photons is modeled as
\begin{align}
    f_z(v_z)=\sqrt{\frac{m\beta_z}{2\pi}}\frac{e^{-\beta_z\frac{m\left(v_z-v_{z,0}\right)^2}{2}}+e^{-\beta_z\frac{m\left(v_z+v_{z,0}\right)^2}{2}}}{2},\label{eq:vzdistr}
\end{align}
where $\beta_z$ is the width determined by the effective temperature reached and $v_{z,0}$ is an offset that models possible misalignment with the $x$-axis. The two averaged Gaussian distributions take into account the two travel directions of cavity-mode photons. It is natural to consider this function in units of Doppler shift $\delta=kv_z$ for which we can use $\tilde{f}(\delta)d\delta=f(v_z)dv_z$ to derive
\begin{align}
    \tilde{f}(\delta)=\frac{1}{\sqrt{2\pi\delta_D^2}}\frac{e^{-\frac{\left(\delta-\delta_0\right)^2}{2\delta_D^2}}+e^{-\frac{\left(\delta+\delta_0\right)^2}{2\delta_D^2}}}{2}.
\end{align}
Here we have defined the Doppler width as
	\begin{align}
	\delta_D=&\sqrt{\frac{k^2}{m\beta_z}}.
	\end{align}

Figure~\ref{Fig:2}(b) shows examples of the velocity distribution along the cavity for $\delta_D=2\pi\times\qty{2}{MHz}$ and $\delta_0=kv_{z,0}=0$ (solid) or $\delta_0=2\pi\times\qty{3}{MHz}$ (dashed). From these curves, we see that $v_{z,0}$ mimics an effective splitting of the velocity distribution.

To prepare the atoms in the inverted state $\ket{e}$, additional experimental steps along the $x$ axis are introduced. With these steps, we aim to realize inversion while controlling the atomic velocity distribution along the cavity axis, to explore different velocity regimes. Particularly intriguing are scenarios where the single-atom linewidth of $\ket{g}\rightarrow\ket{e}$ is Doppler-dominated or transit-time-dominated, as distinct collective behaviors have been predicted for an inverted sample in these two cases \cite{jager2021_1, jager2021_2}.
Although a Doppler-dominated sample is easy to achieve, the transit-time-dominated regime requires slow atoms, along the cavity axis, traversing the cavity mode. This introduces two problems: how does one achieve a variable, narrow velocity distribution and how does one remove the 'undesired' atoms, i.e. those that are too fast.
To solve these problems, we designed a velocity-selective pumping scheme along the $z$ axis that shields the slow atoms by storing them in the metastable states $\ket{m} = {}^3P_{0, 2}$. Subsequently, a cleaning mechanism is employed to accelerate atoms that remain in the ground state because of imperfect state transfer. This is done by Doppler displacing their frequency, to spectrally separate them from the selected atoms.
The velocity selection is achieved by optical pumping on the~\qty{7.5}{kHz}-wide ${}^1S_0\rightarrow {}^3P_1$ transition. We can regulate the mean atomic velocity by choosing the beam angle and frequency detuning, while the velocity range is regulated by the power-broadening of the transition. 
We use resonant light that targets a velocity distribution centered at zero. Tuning the \qty{689}{nm} pump beam to a maximum intensity of~\qty{44.3}{mW/cm^2} ($\sim 10^4I_{\rm sat}$, with $I_{\rm sat} \sim \qty{3}{\mu W/cm^2}$ for this transition), we can address atoms in a range of up to $\pm$~\qty{0.5}{m/s}, giving a maximum Doppler width of~$2\pi\times\qty{0.7}{MHz}$. Here we are limited by the available optical power, but the width of the target distribution can be additionally broadened by introducing a divergence of the storage beam.
The driven atoms are then transferred to ${}^3S_1$ with a~\qty{688}{nm} beam from where they decay to the metastable states. The long lifetimes of these states allow us to store atoms in them for the duration of their travel through the vacuum chamber. The maximum transfer efficiency that we achieve from the full ground state distribution to the storage states, $\ket{m}$, is 67\% for a nozzle temperature of~\qty{803}{K}.

\subsubsection{Sources of decoherence: fast atoms and scattered blue light} \label{sssec:2.1C}
The storage stage selects atoms whose transit-time broadened spectra dominate over their Doppler detuning, that is, atoms for which the condition $v_z \tau < \lambda_{\ket{g}\rightarrow\ket{e}}/2$ ~\cite{Liu:2020} is met. This condition teaches us which atoms contribute to the collective interaction. In fact, for atoms with very small transit times, even high velocities along $z$ do not hinder the interaction with the zero-velocity class. However, after the storage state, atoms that are fast along the cavity-axis remain in the ground state. The presence of these ground-state atoms thus effectively diminishes the inversion and coherence of a collective state.
To address this, we accelerate the remaining ground state atoms using resonant~\qty{461}{nm} light, in a cleaning stage.
The atoms in metastable states do not interact with this light, while ground-state atoms gain a Doppler shift with respect to the bare cavity mode, thereby reducing their coupling strength. 
With 30\,mW of resonant ~\qty{461}{nm} light we reach $\sim 3 I_{\rm sat}$ for a maximum Doppler shift of $2\pi \times$ 7.7\,MHz.  

A second source of decoherence is the blue stray light from 461-nm beams that scatters off atoms and subsequently crosses the cavity mode. 
This blue stray light contributes to the $T_2$ dephasing with an incoherent scattering rate $\gamma_B$. We measure the direct scattering from the cavity atoms and infer a maximum scattering rate $\gamma_B =2\pi\times\qty{9.3}{kHz}$ off atoms in the cavity mode. This rate is comparable to the natural linewidth of the transition and much smaller than $\delta_D$, $1/\tau$, and $\kappa$, which justifies the neglect of this term in our theoretical description in Sec.~\ref{sec:3}.

\subsubsection{\label{sssec:2.1D}Incoherent pumping to $\ket{e}$}
After the velocity-selective storage and state cleaning regions, the atomic beam passes through a heat shield that partially isolates the cavity mode from the first two stages. This is primarily used to reduce heat transfer through thermal radiation from the oven, which can induce a drift of the cavity length when the oven temperature changes.
After the heat shield, the atoms reach the excitation stage. Here, the atoms are pumped with 679-nm and 707-nm $\sigma^\pm$ laser beams propagating along the $y$-axis, with an elliptical beam profile (Gaussian profile along $x$, with a waist of $w_x = 1$~mm and a uniform profile along $z$ with a width of $w_z = 15$~mm). These beams excite the atoms from the metastable states ${}^3P_0$ and ${}^3P_2$, where they are stored, to ${}^3S_1$. From this state, the atoms decay spontaneously via the only remaining decay channel to ${}^3P_1$, see Fig.~\ref{Fig:2}(b). To maximize the gain in our setup we select a single $m_j$ state, the magnetic insensitive $m_j=0$ state, as the excited state. However, the spontaneous decay process does not selectively transfer atoms to this state, so we apply an additional 688-nm $\pi$ polarized beam to depopulate the $m_j = \pm 1 $ states and bring them back into the pumping cycle.\\
The atomic state decays over a few millimeters due to the natural lifetime of the ${}^3P_1$ manifold, making the distance from this stage to the cavity mode crucial. For this reason, the pumping stage is applied as close as possible to the cavity mode \cite{Tang:2022}.

To study the atom-cavity interaction for ground-, metastable- or excited-state atoms, we directly monitor the transmission through the cavity when applying only cooling ($\ket{g}$), adding storage and cleaning ($\ket{m}$), or using all three stages ($\ket{e}$).

\subsubsection{Output detection}\label{sssec:2.1E}
By continuously observing the transmission of an external drive coupled to the TEM$_{00}$-mode of the cavity, we can assess the state of our atomic beam as it traverses the cavity mode. In particular, we can measure the effective atom number coupling to the optical mode, the atomic velocity distribution along $z$ and the inversion efficiency, see Sec.~\ref{sec:3}.
For nominal intracavity intensities of $I_0 = 10^{3}$~to~$10^{4} I_{\mathrm{sat}}$, corresponding to hundreds of $nW$~to~$\upmu W$ input power, we use an avalanche detector to perform a direct measurement of the signal $I_{\Theta}$. Here we define $I_{\Theta}$ as the intracavity intensity modified by atoms in the ground, metastable, or excited state corresponding to $\Theta \in \{g, m, e\}$.
To detect small intensities, down to $I_0 \lesssim I_{\mathrm{sat}}$,  we employ a heterodyne beat measurement, with a $20$~MHz detuned local oscillator (LO). 
For each experimental condition, we can measure a transmission signal proportional to the intracavity field intensity. In particular, the signal $I_{m}$, is used to mimic an empty-cavity signal, even at high oven temperatures. This allows fast modulation between signals $\propto I_{g}$, $I_{m}$ and $I_{e}$, which is convenient, e.g., for fast optimization of atomic state-dependent quantities, by looking at the relative difference with and without atoms at each temperature (see Fig.~\ref{Fig:1}(c) for examples).

\section{\label{sec:3}Theoretical description}
In this section, we present the theoretical tools that we employ to describe our experiment. For this we build on the theory that has been developed in Refs.~\cite{Liu:2020,jager2021_1,jager2021_2,jager2021_3} to describe superradiant emission from atomic beams. Our starting point are the Heisenberg-Langevin equations describing the dynamics of the transition matrix element $\hat{\sigma}_j^-=\ket{g}_j\bra{e}$, the inversion $\hat{\sigma}_j^z=\ket{e}_j\bra{e}-\ket{g}_j\bra{g}$, the atomic position $\hat{x}_j$ and conjugate momentum $\hat{p}_j$ and the cavity annihilation operator $\hat{a}$. The Heisenberg-Langevin equations are given by
	\begin{align}
	\frac{d\hat{a}}{dt}=&\left(i[\Delta_L-\Delta_c]-\frac{\kappa}{2}\right)\hat{a}-i\frac{\Omega+g_0\hat{J}^-}{2}+\sqrt{\kappa}\hat{\mathcal{F}},\label{eq:cav}\\
	\frac{d\hat{\sigma}_j^-}{dt}=&i\Delta_L\hat{\sigma}_j^-+\frac{ig_0}{2}\eta (\hat{\bf x}_j)\hat{\sigma}^z_j\hat{a},\label{eq:sigma}\\
	\frac{d\hat{\sigma}_j^z}{dt}=&ig_0\eta(\hat{\bf x}_j)\left(\hat{a}^\dag \hat{\sigma}_j^--\hat{\sigma}^+_j\hat{a}\right),\label{eq:sigmaz}\\
	\frac{d\hat{\bf x}_j}{dt}=&\frac{\hat{\bf p}_j}{m}.\label{eq:x}
	\end{align}
Equation~\eqref{eq:cav} describes the dynamics of the cavity coupled to atoms with vacuum Rabi frequency $g_0$ and driven by an external laser with rate $\Omega$. Here, we introduce the detunings $\Delta_L=\omega_L-\omega_a$ between the laser and the atomic transition, and $\Delta_c=\omega_c-\omega_a$ between the cavity and the atomic transition. Furthermore, we introduced the collective dipole
	\begin{align}
	\hat{J}=\sum_{j}\eta(\hat{\bf x}_j)\hat{\sigma}_j,
	\end{align}
which weights the individual atomic transition elements with the cavity mode function evaluated at their current atomic position $\hat{x}_j$. The last term in Eq.~\eqref{eq:cav} describes cavity shot noise with $\langle\hat{\mathcal F}(t)\rangle=0=\langle\hat{\mathcal F}^\dag(t)\hat{\mathcal F}(t')\rangle$ and $\langle\hat{\mathcal F}(t)\hat{\mathcal F}^\dag(t')\rangle=\delta(t-t')$. Equations~\eqref{eq:sigma} and \eqref{eq:sigmaz} describe the dynamics of the transition matrix element and the atomic inversion coupled to the cavity while Eq.~\eqref{eq:x} describes ballistic motion of atoms with mass $m$. We assume that the atoms enter the cavity in an incoherent mixture and traverse the cavity mode profile, which we describe by
	\begin{align}
	\eta({\bf x})=\cos(kz)e^{-\frac{x^2}{w^2}}.\label{eq:eta}
	\end{align}
Here, $w$ is the waist of the cavity mode (see Table~\ref{Table:1}) and $k=2\pi/\lambda$ is the wavenumber of the cavity mode. Note that we restrict ourselves to an effective two-dimensional model that includes the most important concepts such as Doppler-broadening, given by
	\begin{align}
	\delta_D=\frac{k\Delta \hat{p}_z}{m}\label{eq:doppler1}
	\end{align}
	and transit time
	\begin{align}
	\tau=\frac{2w}{\langle \hat{p}_x\rangle /m},\label{eq:transit1}
	\end{align}
where $\Delta p_z=\sqrt{\langle \hat{p}_z^2\rangle-\langle\hat{p}_z\rangle^2}$ is the width of the single-particle momentum distribution along the cavity axis and $\langle \hat{p}_x\rangle/m$ the mean velocity along the atomic beam axis. 
	
Equations~\eqref{eq:sigma}-\eqref{eq:x} are derived assuming that (i) spontaneous emission and external sources of dephasing are negligible on the typical timescale in which the atoms interact with the cavity. As we discussed in the previous section this is fulfilled since our typical interaction time is determined by the transit time of the atoms, which is $\lesssim1\mu$s, while the decoherence times from free-space spontaneous emission and scattered blue light are of the order $\sim100\mu$s. In addition, we assume that (ii) the atoms transit the cavity ballistically and optomechanical forces are negligible. This is true if the typical exchange of momentum does not significantly modify the atomic momentum distribution,  $\hbar k g\tau  \sqrt{\langle\hat{a}^\dag\hat{a}\rangle}\ll \Delta p$, where $\langle\hat{a}^\dag\hat{a}\rangle$ denotes the mean intracavity photon number.

\subsection{Mean-field description}
To be able to simulate the described equations and to extract analytical results we employ a mean-field description of the atoms. Here, we exchange the operators by complex numbers $\hat{a}\leftrightarrow\alpha$, $\hat{\sigma}_j^-\leftrightarrow s_j$, $\hat{\sigma}_j^z\leftrightarrow s_j^z$, $\hat{x}_j\leftrightarrow x_j$, and $\hat{p}_j\leftrightarrow p_j$. In addition we neglect the shot noise contribution so that the corresponding mean-field equations read
	\begin{align}
	\frac{d\alpha}{d t}=&\left(i[\Delta_L-\Delta_c]-\frac{\kappa}{2}\right)\alpha-i\frac{\Omega+gJ}{2}\label{eq:alpha}\\
	\frac{ds_j}{dt}=&i\Delta_L s_j+\frac{ig}{2}\eta ({\bf x}_j)s^z_j\alpha,\label{eq:meansj}\\
	\frac{ds_j^z}{dt}=&ig\eta({\bf x}_j)\left(\alpha^*s_j-s_j^*\alpha\right),\label{eq:meansjz}\\
	\frac{d{\bf x}_j}{dt}=&\frac{{\bf p}_j}{m},\label{eq:meanx}
	\end{align}
with $J=\sum_{j}\eta({\bf x}_j)s_j$. When performing numerical simulations, we use these equations of motion (for further details, we refer to Appendix~\ref{App:Numerical}). For analytical calculations it is useful to introduce  a description for the spin densities $s({\bf x},{\bf p},t)=\sum_js_j\delta({\bf x}-{\bf x}_j)\delta({\bf p}-{\bf p}_j)$ and $s^z({\bf x},{\bf p},t)=\sum_js_j^z\delta({\bf x}-{\bf x}_j)\delta({\bf p}-{\bf p}_j)$ given by
	\begin{align}
	\frac{\partial s}{\partial t}=&-\frac{\bf p}{m}\cdot\nabla_{\bf x}s+i\Delta_L s+\frac{ig}{2}\eta ({\bf x})s^z\alpha,\label{eq:mfs}\\
	\frac{\partial s^{z}}{\partial t}=&-\frac{\bf p}{m}\cdot\nabla_{\bf x}s^z+ig\eta({\bf x})\left(\alpha^*s-s^*\alpha\right).\label{eq:mfz}
	\end{align}
The collective dipole in the density description can be calculated as
	\begin{align}
	J=\int d{\bf x}\int d{\bf p}\eta({\bf x})s({\bf x},{\bf p},t).\label{eq:J}
	\end{align}
Above, we used the notation $\nabla_{\bf x}=(\partial/[\partial x],\partial/[\partial z])^T$ and as boundary conditions we use $s({\bf x}_0,{\bf p})=0$ and $s^z({\bf x}_0,{\bf p})=s_0^z({\bf p})$ where ${\bf x}_0=(x_0,z_0)$ is an arbitrary position on the line where $x_0\rightarrow-\infty$. We assume that $s_0^z({\bf p})$ is homogeneous in space, which assumes that the atomic beam width is much larger than the wavelength of the mode function $\cos(kz)$. This also enables us to restrict the spatial extent of the integration $\int d{\bf x}=\int_{0}^\lambda dz\int_{-\infty}^{\infty} dx$ while the integration over momentum is $\int d{\bf p}=\int_{0}^\infty dp_x\int_{-\infty}^\infty dp_z$. To model the physical state of the atomic beam, we further assume
	\begin{align}
	s_0^z({\bf p})=\frac{N_ef_x^{(e)}\left(\frac{p_x}{m}\right)f_z^{(e)}\left(\frac{p_z}{m}\right)-N_gf_x^{(g)}\left(\frac{p_x}{m}\right)f_z^{(g)}\left(\frac{p_z}{m}\right)}{2w\lambda m^2}.\label{eq:s0z}
	\end{align}
 The numbers $N_g$ and $N_e$ describe the number of atoms in the excited and ground state within an interval of $x\in[-w,w]$. Note that this defines the number of atoms coupled to the cavity as $N=N_e+N_g$. The terms $f_x^{(a)},f_z^{(a)}$ describe the momentum distribution of the atoms, which has been introduced in Eqs.~\eqref{eq:vxdistr} and \eqref{eq:vzdistr} where we allow for different values of Doppler width $\delta_D^{(a)}$ and offset $\delta_0^{(a)}$ for atoms in ground and excited states, $a=g$ and $a=e$, respectively. The physical reason for the dependence of the Doppler width and offset is based on our preparation protocol. The alignment of the cooling, cleaning, collimation, and excitation lasers to the cavity axis all affect the final atomic distribution. 
In conclusion, we assume that this distribution models our physical reality at hand, although we note that one could improve the modeling by incorporating a correlation between $p_x$ and $p_z$. In this paper we will, however, assume no such correlation and additionally that the excited and ground state atoms move the same way along the $x$ axis resulting in $f_x^{(e)}=f_x^{(a)}$.
 
In the following, we discuss the stationary state of Eqs.~\eqref{eq:alpha}, \eqref{eq:mfs}, and \eqref{eq:mfz}. The first parameter regime that we consider is the weak driving regime, where the change in the atomic states can be treated perturbatively. This will enable us to derive the condition for lasing and normal mode splitting analytically.

\subsection{Weak driving regime}
In the weak driving regime we assume $\Omega$, $\alpha=\delta\alpha$ and $s=\delta s$ are small so that we can neglect any additional effects on the internal state population $s_0^z$. The formal solution for $\delta s$ is then given by
	\begin{align}
	\delta s=\frac{ig\delta\alpha}{2}\left[-i\Delta_L+\frac{\bf p}{m}\cdot \nabla_{\bf x}\right]^{-1}\eta({\bf x})s_0^z.
	\end{align}
	With this solution we can find an expression for the collective dipole [Eq.~\eqref{eq:J}] given by
	\begin{align}
	\delta J=&\int d{\bf x}\int d{\bf p}\eta({\bf x})\delta s\nonumber\\
	=&\frac{ig\delta\alpha}{2}\int d{\bf x}\int d{\bf p}\int_{0}^{\infty}dt\,e^{i\Delta_L t}\eta\left({\bf x}+\frac{\bf p}{m}t\right)\eta({\bf x})s_0^z.
	\end{align}
	Now solving Eq.~\eqref{eq:alpha} for the field, we get
	\begin{align}
	\delta\alpha=&\frac{-i\frac{\Omega}{2}}{-i[\Delta_L-\Delta_c]+\frac{\kappa}{2}+\frac{\Gamma}{2}},\label{eq:deltaalpha}
	\end{align}
	where we introduced
	\begin{align}
	\Gamma=-\frac{g^2}{2}\int d{\bf x}\int d{\bf p}\int_{0}^{\infty}dt\,e^{i\Delta_L t}\eta\left({\bf x}+\frac{\bf p}{m}t\right)\eta({\bf x})s_0^z.\label{eq:Gamma}
	\end{align}
	This allows us to calculate the normalized transmission
	\begin{align}
	\frac{I}{I_0}=\frac{|\alpha|^2}{(\Omega/\kappa)^2}.\label{eq:normalizedI}
	\end{align}
We can now use Eq.~\eqref{eq:s0z} to derive
	\begin{align}
	\Gamma=&\Gamma_g-\Gamma_e\label{eq:Gamma2}
	\end{align}
 with
 \begin{align}
 \Gamma_a=&\sqrt{\frac{\pi}{2}}\frac{N_ag^2}{8}\int_{0}^\infty dt \, e^{i\Delta_Lt} \cos(\delta_0^{a}t)e^{-\frac{(\delta_D^{(a)})^2t^2}{2}}\frac{1}{1+\frac{8t^2}{\pi\tau^2}},
 \end{align}
 and $a=e,g$. The rates $\Gamma_g$ and $\Gamma_e$ describe the loss and gain resulting from the presence of atoms in the ground and excited states, respectively. If all atoms are in the ground state, then $N_e=0$ and the presence of the atoms creates an additional loss channel.  For the extreme case $N_g=0$, $N_e=N$ and we obtain $\Gamma=-\Gamma_G<0$ which results in gain for the cavity field mode.

In order to understand the physical implications of this formula, we study the case of the cavity being resonant with the atoms, $\Delta_c=0$, which is also considered in the experiment.	

\subsubsection{Absorption and normal mode splitting}
We first study $N_e=0$, which is the physical scenario of absorption. For this scenario, we set $\delta_0=\delta_0^{(g)}$ and $\delta_D=\delta_D^{(g)}$. The theoretical results discussed in this part are important for the fitting procedure that is used in Sec.~\ref{sec:4:1} and for Figs.~\ref{Fig:6} and \ref{Fig:7}.

\begin{figure}
\centering
	\includegraphics[width=1\linewidth]{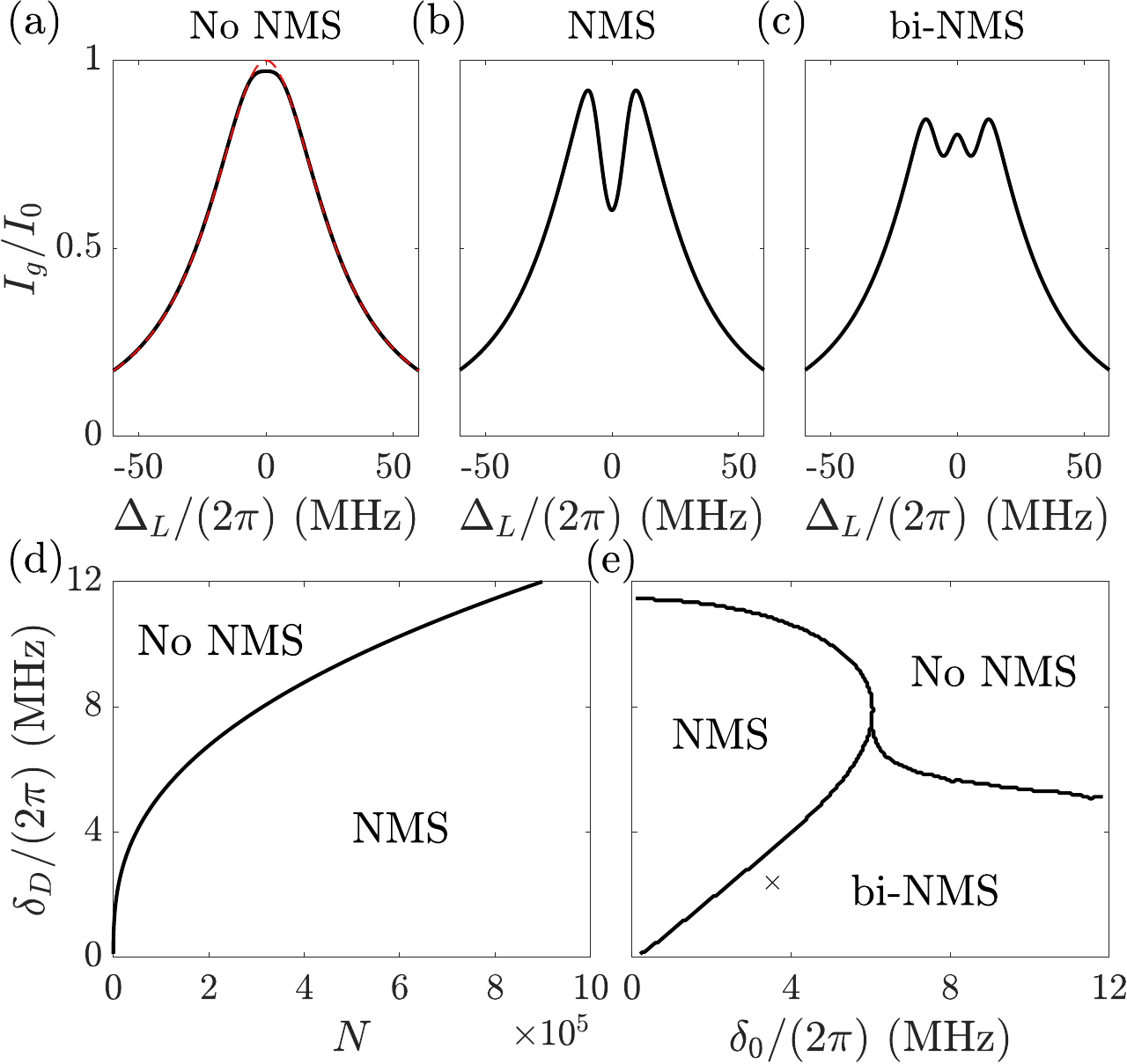}
	\caption{\label{Fig:3}Transmission as function of $\Delta_L$ calculated using Eqs.~\eqref{eq:normalizedI}, \eqref{eq:Gamma2}, and Eq.~\eqref{eq:deltaalpha} for (a) $N=4\times10^4$, $\delta_0=0$, $\delta_D=2\pi\times\qty{4}{MHz}$, (b) $N=8\times10^5$, $\delta_0=0$, $\delta_D=2\pi\times\qty{4}{MHz}$, and (c) $N=8\times10^5$, $\delta_0=2\pi\times\qty{5.5}{MHz}$, $\delta_D=2\pi\times\qty{4}{MHz}$. (d) Phase diagram indicating the regions of no normal mode splitting [No NMS, see (a)] and normal mode splitting [NMS, see (b)] as function of $\delta_D$ and $N$ for fixed $\delta_0=0$. (e) Phase diagram showing the parameter space in $\delta_0$ and $\delta_D$ where one observes no normal mode splitting [No NMS, see (a)],  normal mode splitting [NMS, see (b)] and binary normal mode splitting [bi-NMS, see (c)] for fixed $N=8\times10^5$. The cross symbol in (e) indicates the parameters that we realized experimentally (see Fig.~\ref{Fig:7}) . The remaining parameters are $\kappa=2\pi\times\qty{54}{MHz}$, $\tau=\qty{0.5}{\upmu s}.$}
\end{figure}

We discuss the case where the atoms in the atomic beam can absorb cavity photons and leave the cavity with the resulting excitation. This creates an additional decay channel from the cavity and results in an effective broadening of the cavity linewidth. In the formulas, this is expressed by $\Gamma>0$ for $\Delta_L=0$. In Fig.~\ref{Fig:3}(a) we plot the transmission of the cavity for $N=10^4$ and $\delta_0=0$. We see a transmission that is only very slightly modified with respect to the empty cavity. To demonstrate the broadening, we plot the empty cavity transmission for $N=0$ as a red dashed line. In particular, we see an absorption close to $\Delta_L\approx 0$. When increasing the atom number we observe a normal mode splitting (NMS), which is visible in Fig.~\ref{Fig:3}(b) for $N=8\times 10^5$ and $\delta_0=0$.
To study the onset of NMS, we show in Fig.~\ref{Fig:3}(d) a phase diagram where we indicate the parameters of $\delta_D$ and $N$ where no-NMS and NMS are seen, respectively. The phase diagram is calculated for $\delta_0=0$. We see that NMS requires sufficiently large atom numbers and sufficiently low Doppler width $\delta_D$. For realistic experimental parameters of $\delta_D\sim\qty{5}{MHz}$ we thus expect a NMS if $N>10^5$.

To explore the role of $\delta_0$ we show in Fig.~\ref{Fig:3}(c) the transmission for $\delta_0=2\pi\times\qty{5.5}{MHz}$ and $N=8\times 10^5$. In contrast to Fig.~\ref{Fig:3}(b) we now see three peaks with a symmetry around $\Delta_L=0$. This "binary NMS" (bi-NMS) is a consequence of two normal mode splittings that appear at the offset resonance $\pm \delta_0$, this leads to a total of three peaks. To show the parameter regime where this behavior is observed, we calculated in  Fig.~\ref{Fig:3}(e) the phase diagram that shows all three transmission regimes. The phase diagram is plotted for fixed atom number $N=8\times10^5$. For low $\delta_0$ we observe mostly NMS, which can, however, transition to no-NMS for sufficiently large Doppler width $\delta_D$. Increasing $\delta_0$ we can enter the bi-NMS regime provided that the offset can overcome the broadening $\delta_0 \gtrsim \delta_D$.

\subsubsection{Gain and Lasing}
\begin{figure}
\centering
	\includegraphics[width=1\linewidth]{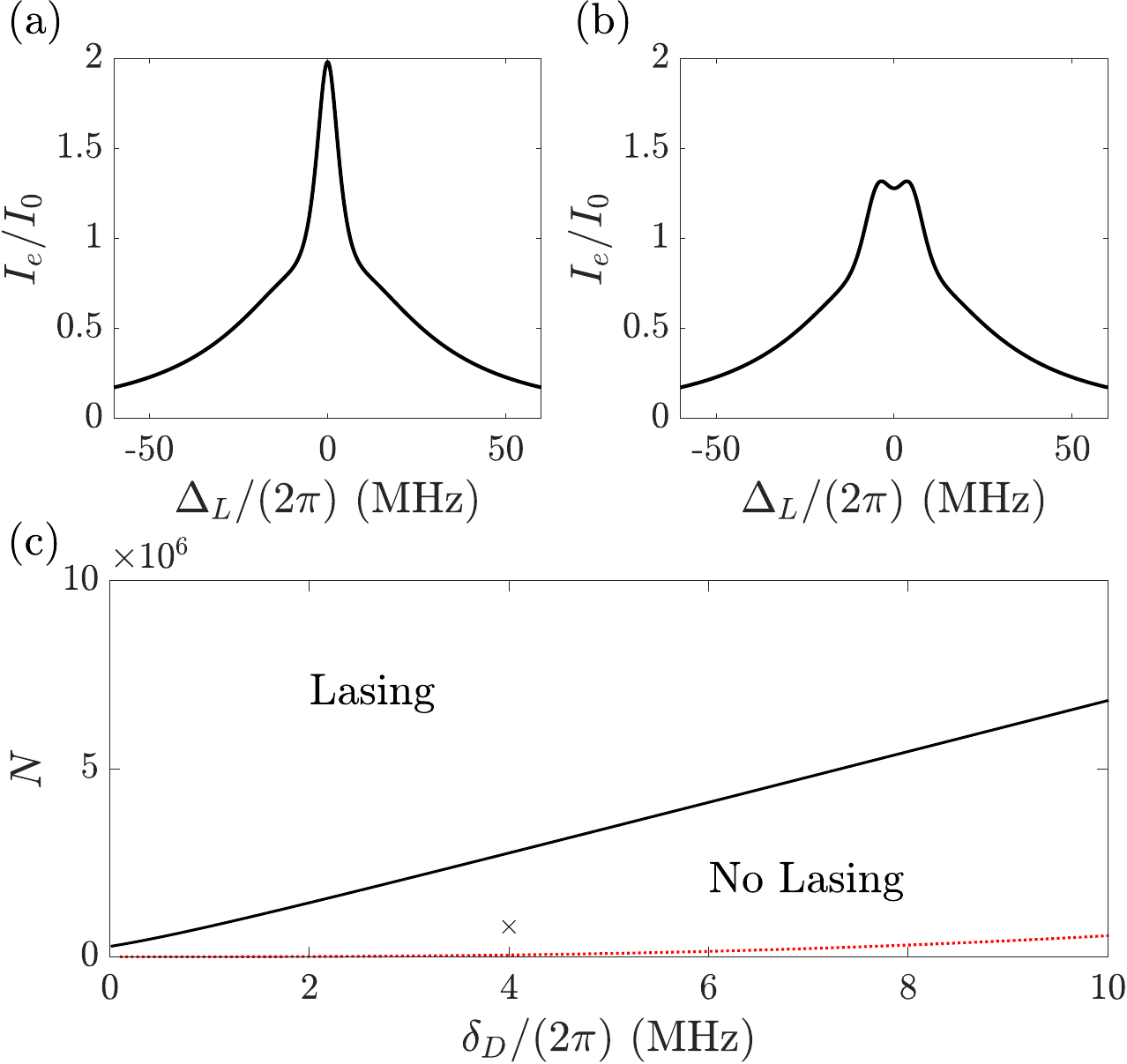}
	\caption{\label{Fig:4}Transmission as function of $\Delta_L$ calculated using Eqs.~\eqref{eq:normalizedI}, \eqref{eq:Gamma2}, and Eq.~\eqref{eq:deltaalpha} for $N=8\times10^5$ and (a) $\delta_0=0$, $\delta_D=2\pi\times\qty{4}{MHz}$, (b) $\delta_0=2\pi\times\qty{5.5}{MHz}$, $\delta_D=2\pi\times\qty{4}{MHz}$. (c) Threshold to superradiant lasing for $\delta_0=0$ as function of $N$ and Doppler-width $\delta_D$. In all calculations we have assumed a perfectly inverted atomic beam. The red dashed line shows the threshold above which NMS would be found when $N=N_g$. The remaining parameters are $\kappa=2\pi\times\qty{54}{MHz}$, $\tau=\qty{0.5}{\mu s}.$ The cross marker indicates the parameters used in subplot (a) and it points to the experimentally achieved numbers of $N$ and $\delta_D$ before atomic excitation in our setup (see Sec.~\ref{sec:4} and Fig.~\ref{Fig:7}).}
\end{figure}

We will now study the regime in which we expect a gain from an atomic inversion that can result in lasing. For this, we focus on the case $N_g=0$ and $N_e=N$. In this case, we assume $\delta_0=\delta_0^{(e)}$ and $\delta_D=\delta_D^{(e)}$. In Fig.~\ref{Fig:4}(a) we show a typical transmission profile if the atoms are perfectly inverted for $N=8\times10^5$, $\delta_0=0$, and a Doppler width of $\delta_D=2\pi\times\qty{4}{MHz}$. For these parameters, we already see a large gain at resonance $\Delta_L\approx0$. Increasing the atom number further will eventually result in lasing. We can calculate the lasing threshold by finding the divergence of the transmission originating from a divergence of Eq.~\eqref{eq:deltaalpha}. The solid line in Fig.~\ref{Fig:4}(c) shows the transition from non-lasing to lasing where we have chosen $\delta_0=0$. For the experimentally relevant parameters, we find that lasing requires atom numbers of several million. In the experiment, we reach an order of magnitude less in atom number. As a guide to the eye, we also show the threshold for NMS as a red dashed line. The transition to NMS occurs for atom numbers that are an order of magnitude lower, which highlights that the presence of NMS does not imply that one can access a lasing regime. 

To demonstrate the effect of a velocity offset described by $\delta_0\neq0$ we show the cavity transmission in Fig.~\ref{Fig:4}(b) for $\delta_0=2\pi\times\qty{5.5}{MHz}$. Here, the maximum gain is found at a different location, $\Delta_L\approx\delta_0$, highlighting the overall Doppler-shifted emission frequency of the gain medium. We remark at this point that this transmission profile is a precursor of the ``bistable superradiant'' phase described in Ref.~\cite{jager2021_1}. This is a distinct superradiant lasing phase, which can in principle be accessed in such setups. This gain profile is qualitatively recovered by our experiment as discussed in Sec.~\ref{sec:4:2} (see also Fig.~\ref{Fig:9}).

\subsection{Stronger driving}
So far, we have only discussed the implications of weak cavity probing. In this subsection, we show the effects of strong driving in the transmission profile and also derive a semi-analytical formula for calculating the transmission on resonance. This formula is used for the experiment comparison discussed in Sec.~\ref{sec:4:2} and Fig.~\ref{Fig:8}.

\begin{figure}
\centering
	\includegraphics[width=1\linewidth]{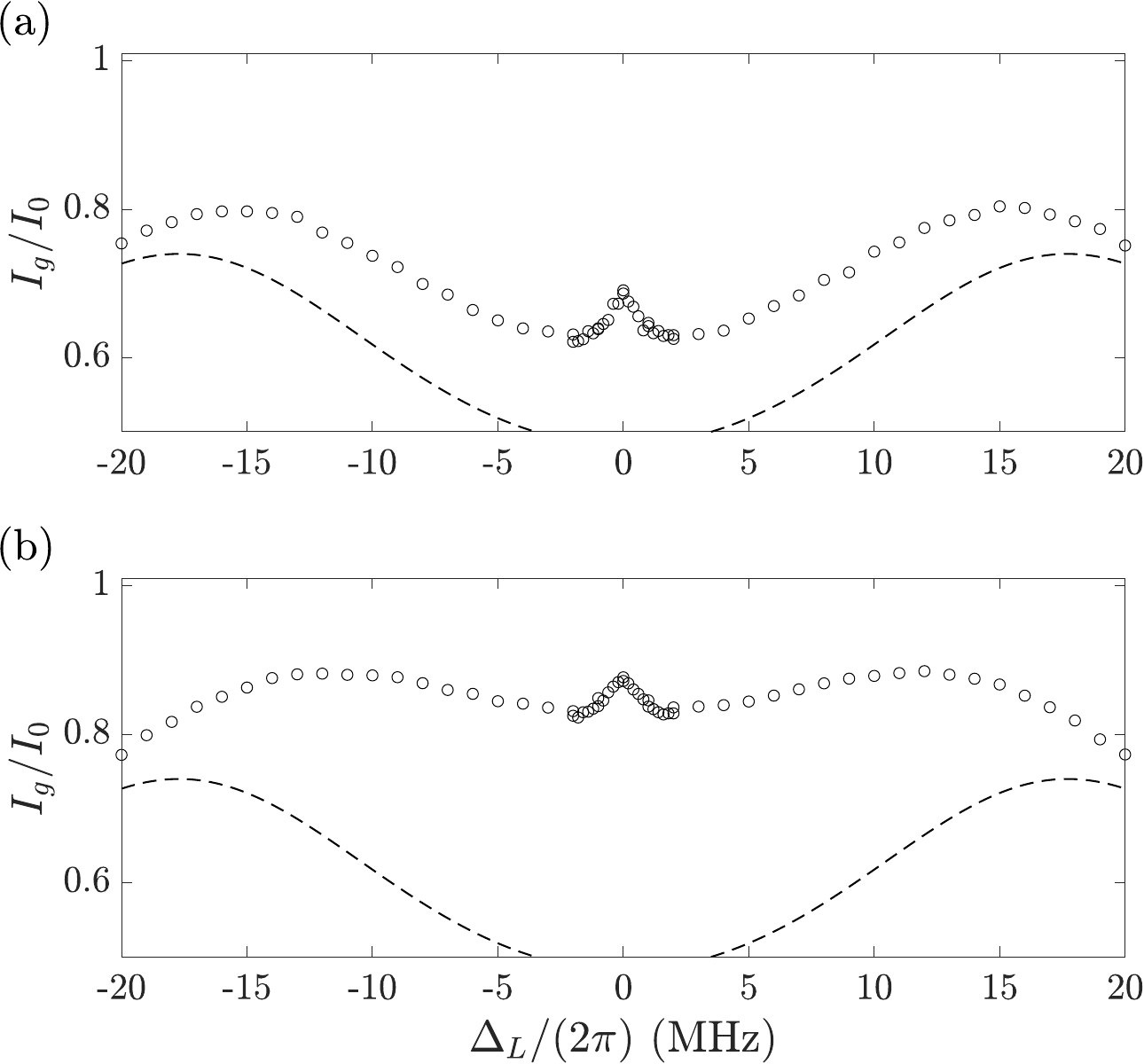}
	\caption{\label{Fig:5}Transmission as function of $\Delta_L$ for $N=3\times10^6$, $\delta_D=2\pi\times\qty{10}{MHz}$, $\delta_0=0$ and (a) $I_0/I_{\mathrm{sat}}=9.20\times10^4$ and (b) $I_0/I_{\mathrm{sat}}=3.68\times10^5$, respectively. The dashed lines are the prediction of the weak driving theory using Eq.~\eqref{eq:deltaalpha}. For the calculation of the data points we have simulated the mean-field equations~\eqref{eq:alpha}-\eqref{eq:meanx}.}
\end{figure}

To access the transmission spectrum, we simulate the mean-field equations~\eqref{eq:alpha}-\eqref{eq:meanx}. We initialize the system in the atomic ground state $N_e=0$ and $N_g=N=3\times10^6$. We work in a regime where we expect NMS for weak driving $\delta_D=\delta_D^{(g)}=2\pi\times\qty{10}{MHz}$, $\delta_0=\delta_0^{(g)}=0$.  In Fig.~\ref{Fig:5}(a) and (b) we show as a dashed line the transmission profile for weak driving calculated using Eq.~\eqref{eq:deltaalpha}. In the simulations we choose parameters where the atoms start to saturate and thus expect discrepancies from Eq.~\eqref{eq:deltaalpha}. In order to classify the input power, we refer to the saturation intensity inside of the cavity, 
\begin{align}
\frac{|\Omega|^2g^2}{\kappa^2\gamma^2}=\frac{I_0}{2I_{\mathrm{sat}}},
\end{align}
where $\gamma=2\pi\times~\qty{7.5}{kHz}$ is the natural linewidth of the atoms.  We remark here that even $I/I_{\mathrm{sat}}\gg1$ does not imply saturation of the atoms, since they only interact with the cavity over a very short timescale $\tau\ll\gamma^{-1}$. In Fig.~\ref{Fig:5}(a) and (b) we show the transmission for very large intensities $I/I_{\mathrm{sat}}=9.20\times10^4$ and $I/I_{\mathrm{sat}}=3.68\times10^5$, respectively. 

The central peak shows up as stationary atoms along the cavity axis are saturated by light propagating in both directions within the cavity. For trapped atomic ensembles, this peak is typically limited by the natural linewidth of the atoms \cite{Rivero:2023, Christensen:2015}. In vapor cells, Doppler broadening determines the width \cite{gea2008}. In our atomic beam setup, the limitation is primarily given by the inverse of the typical interaction time $1/\tau$ of the atoms with the cavity. In particular, we only observe small changes in the width for the different driving strengths used in Fig.~\ref{Fig:5}(a) and (b).

Here we provide the general formula for describing the transmission for strong driving in the specific case of $\Delta_L=0$. We derive a semi-analytical formula for the transmission, which is later used in Fig.~\ref{Fig:8}. For this, we employ the parameterization $s=s_0^z({\bf p})\sin[K({\bf x},{\bf p})]$ and $s^z=s_0^z({\bf p})\cos[K({\bf x},{\bf p})]$, where $K$ is the azimuthal angle and $s_0^z$ describes the state of the incoming atoms at $x_0$. The boundary condition is given by $K({\bf x}_0,{\bf p})=0$. Under these assumption we can derive a differential equation for $K$ that is valid at steady state
	\begin{align}
	\frac{\bf p}{m}\cdot\nabla_{\bf x}K=g\alpha_0\eta({\bf x}),\\
	\alpha_0=\frac{\Omega+gJ}{\kappa}.\label{eq:alpha0}
	\end{align}
	This equation can be solved using the form of $\eta$ given in Eq.~\eqref{eq:eta} and $K({\bf x},{\bf p})$ takes the form
	\begin{align}
	K=&g\alpha_0\int_{-\infty}^{\frac{mx}{p_x}} \,dt'\,\cos\left[kz_0+\frac{kp_z}{m}t'\right]e^{-\frac{p_x^2(t')^2}{m^2w^2}}\nonumber\\
	=&\frac{g\alpha_0m}{p_x}\int_{-\infty}^x\,du\,\cos\left[k z-\frac{kp_z}{p_x}(x-u)\right]e^{-\frac{u^2}{w^2}}.\label{eq:K}
	\end{align}
Now using Eq.~\eqref{eq:K} in Eq.~\eqref{eq:J} we get
	\begin{align*}
	J=&-\int d{\bf x}\int d{\bf p}\frac{s_0^z({\bf p})}{2g\alpha_0}\frac{\bf p}{m}\cdot\nabla_{\bf x}\cos[K({\bf x},{\bf p})]\\
	=&-\int d{\bf x}\int d{\bf p}\frac{s_0^z({\bf p})}{2g\alpha_0}\frac{p_x}{m}\frac{\partial \cos[K({\bf x},{\bf p})]}{\partial x},
	\end{align*}
	since $K$ is periodic in $z$. Combining these two results leads to the equation
	\begin{align}
	J=&\lambda\int d{\bf p}\frac{p_x}{m}\frac{s_0^z({\bf p})}{2g\alpha_0}\left[1-\mathcal{J}_0\left(\frac{g\alpha_0m}{p_x}\sqrt{\pi w^2}e^{-\frac{k^2w^2p_z^2}{4p_x^2}}\right)\right].
	\end{align}
	In order to find the steady state of the cavity field, one now needs to solve the above equation and Eq.~\eqref{eq:alpha0} self-consistently using Eq.~\eqref{eq:s0z}. In practice, this is done numerically.

\section{\label{sec:4} Results and Analysis}
In this section we present experimental data and use the theory introduced in the previous section for their analysis. 

\subsection{\label{sec:4:1}Atomic beam properties as a function of oven temperature}
\begin{figure}
\centering
	\includegraphics[width=1\linewidth]{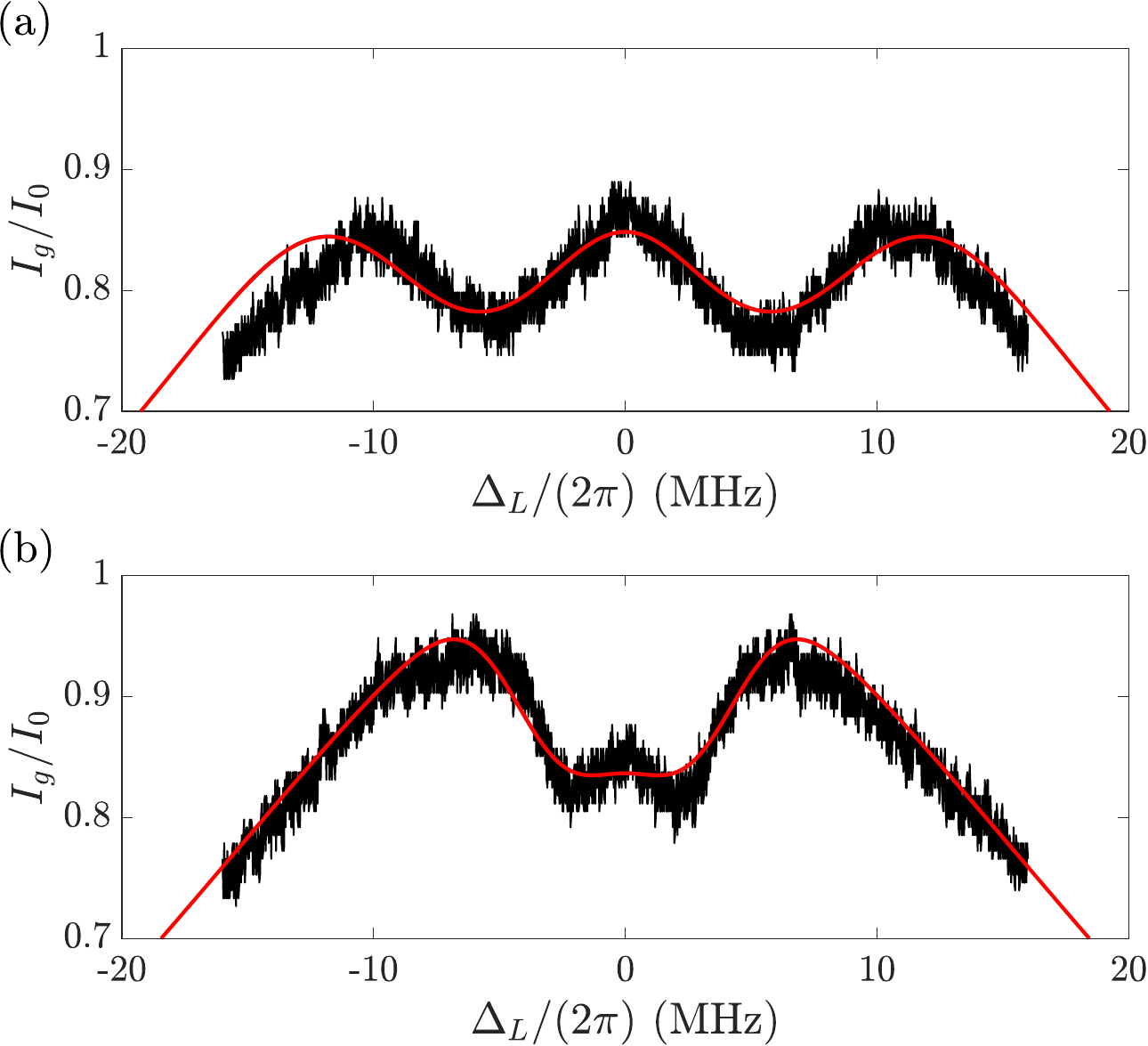}
	\caption{Experimentally measured transmission as function of the detuning $\Delta_L$ for (a) $T=\qty{823}{K}$ and (b) $T=\qty{781}{K}$ with an intra-cavity power corresponding to $I_0/I_{\mathrm{sat}}=1.1\times 10^4$. The red curves are fits to the experimental data of the numerically calculated transmission with optimized $N$, $\delta_D$ and $\delta_0$. For the fits we used Eqs.~\eqref{eq:deltaalpha} and \eqref{eq:Gamma2} (a) $\tau=\qty{0.486}{\mu s}$, (b) $\tau=\qty{0.499}{\mu s}$ and $\kappa=2\pi\times\qty{54}{MHz}$.\label{Fig:6}}
\end{figure}

To study the atomic beam parameters in different regimes, we change the oven temperature while keeping the cooling and collimation protocol the same. Atoms enter the cavity in $\ket{g}$. The bare cavity is resonant with the atomic transition $\ket{g} \rightarrow \ket{e}$. We change the external driving laser detuning $\Delta_L$ over the interval $2\pi\times\left[-\qty{16}{MHz},\qty{16}{MHz}\right]$. The lasers' intensity is fixed and corresponds to a single-atom intra-cavity saturation parameter of $I_0/I_{\mathrm{sat}}=1.1\times 10^4$. 
The results are shown in Fig.~\ref{Fig:6}(a) for $T=\qty{823}{K}$ and (b) for $T=\qty{781}{K}$.
For both temperatures we find a central peak and two shoulders. Given a Doppler-free saturation feature, we would expect the central peak linewidth to be determined by transit time and power broadening. However, in both cases we observe a broader feature. We attribute this behavior to a non-zero $\delta_0=\delta_0^{(g)}$ of the atomic beam, indicating that we are working in the bi-NMS parameter regime. The broadening of the central peak, together with the increased separation of the side shoulders, indicates that by increasing the temperature in Fig.~\ref{Fig:6}(a), we have reached a higher atomic flux and also a larger $\delta_0$. 

\begin{figure}
\centering
	\includegraphics[width=1\linewidth]{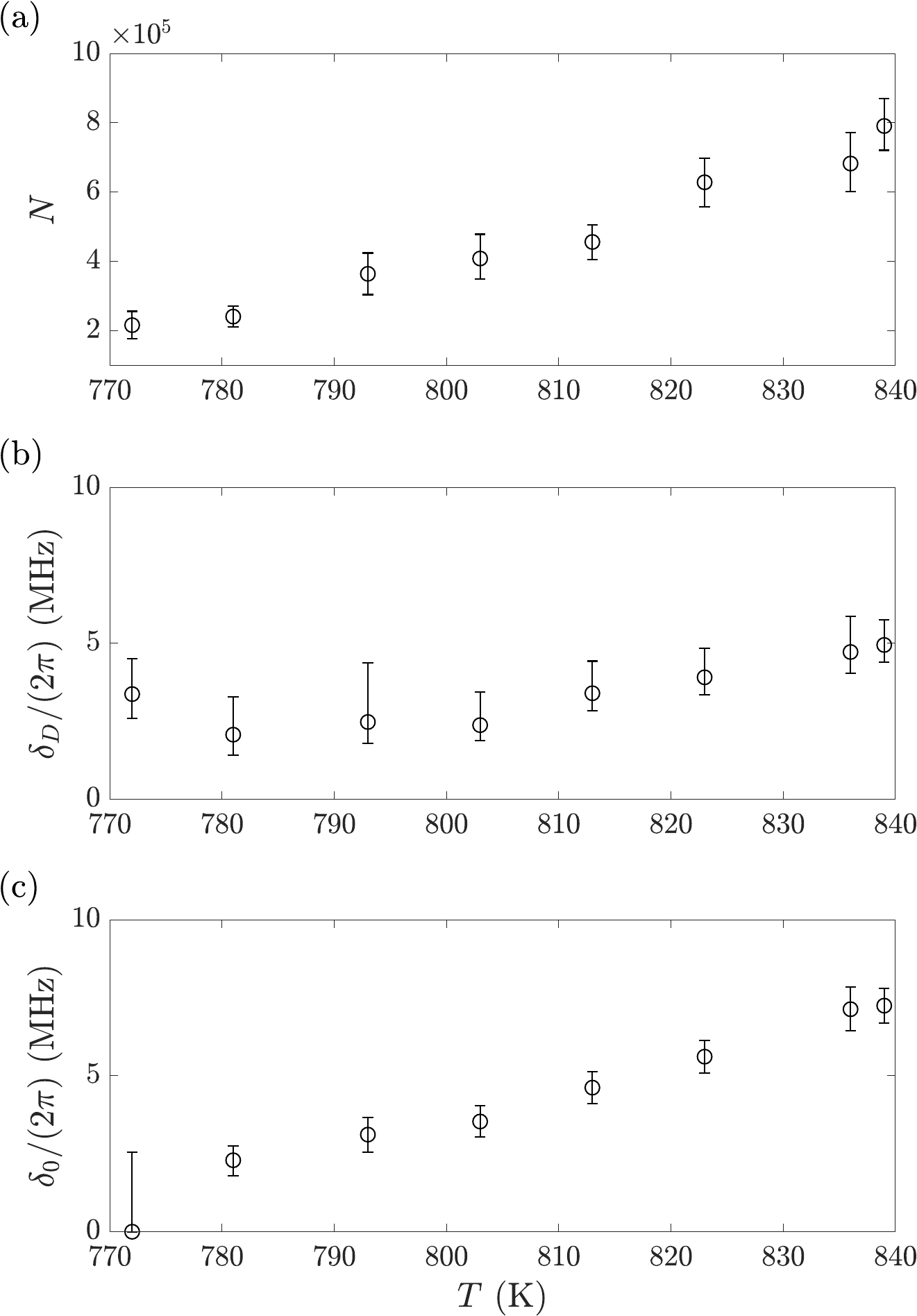}
	\caption{\label{Fig:7}The three parameters (a) $N$, (b) $\delta_D$, and (c) $\delta_0$ obtained by fitting Eq.~\eqref{eq:normalizedI}, using Eqs.~\eqref{eq:deltaalpha} and \eqref{eq:Gamma}, to the measured data for different oven temperatures $T$. The errorbars indicate where the relative error of the fit has increased by more than $10\%$ when changing the single corresponding parameter. For examples of the fits see Fig.~\ref{Fig:6}.  }
\end{figure}

We use Eqs.~\eqref{eq:deltaalpha} and \eqref{eq:Gamma2}, valid for transmission under weak driving, and fit the values of $\delta_D$, $\delta_0$, and $N$ to the experimental data. The corresponding fits are visible as red lines in Fig.~\ref{Fig:6}(a) and (b) and show excellent qualitative and good quantitative agreement. The fitted parameters, including error bars, are shown in Fig.~\ref{Fig:7} against the oven temperature $T$.
In Fig.~\ref{Fig:7}(a) we show the increasing atom number as a function of oven temperature. This implies that our oven operates in a regime where temperature is approximately linearly proportional to the atom number arriving at the cavity. The expected oven vapor pressure for strontium increases by an order of magnitude over this temperature range, indicating that we could be at the edge of the molecular flow regime of the oven nozzle~\cite{beijerinck1975}. 
Figure~\ref{Fig:7}(b) shows that at low temperatures, the cooling protocol consistently prepares the atoms near a Doppler width of approximately $\delta_D=\delta_D^{(g)}\sim2\pi\times\qty{3}{MHz}$ along the cavity axis. With increasing temperature, however, the rising optical density of the atomic beam reduces the effectiveness of the transverse cooling, leading to an increased Doppler-spread. In Fig.~\ref{Fig:7}(c) we show the fitted mean velocity offset $\delta_0$. Notably, we observe a growing offset $\delta_0$ with increasing temperature, which we attribute to a combination of two factors. Firstly, a structural tilt which is independent of temperature. Secondly, imperfections in our transverse cooling caused by spatial and velocity dependence of the absorption in the transverse cooling beams. Since the efficiency of transverse cooling changes with atomic density, we also see a $\delta_D$ dependence of the oven temperature. In short, the optical density of a high-flux beam is problematic for a technique that relies on finely balancing forces. We note that the observed absorption of cooling beams in the experiment from a single pass is on the order of 40~$\%$. The combination of a modified cooling efficiency and the structural tilt produces the temperature dependent offset.

\subsection{Absorption and Gain on resonance\label{sec:4:2}}
\begin{figure}
\centering
	\includegraphics[width=1\linewidth]{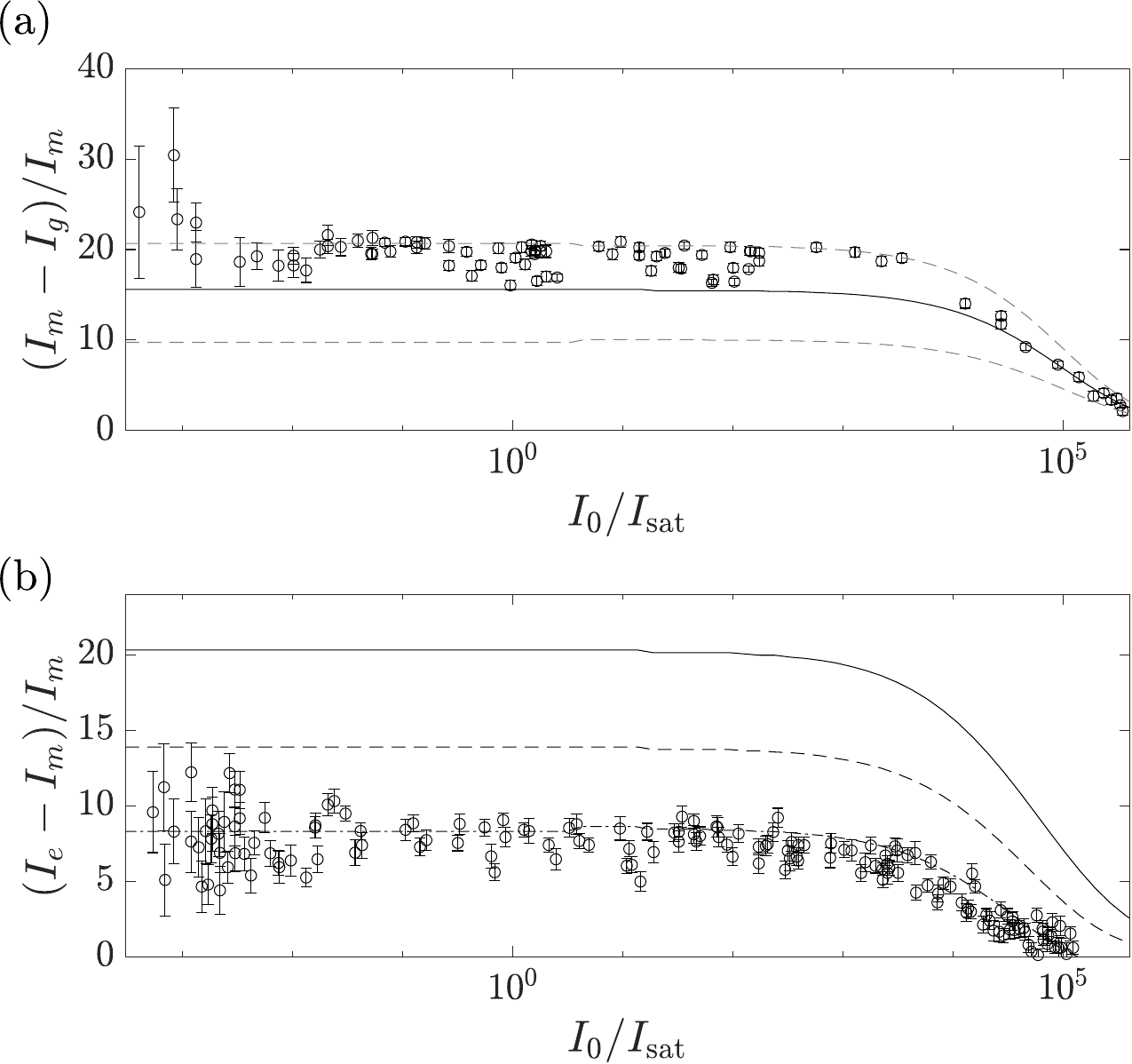}
	\caption{\label{Fig:8}Measured (a) relative absorption and (b) relative gain as function of the input power visible as symbols with measured error bars. The oven temperature is $T=803$~K and we use the fit parameters from Fig.~\ref{Fig:7}, $N=4.1\times10^5$, $\delta_D=2\pi\times\qty{2.38}{MHz}$, and $\delta_0=2\pi\times\qty{3.53}{MHz}$. By solving Eq.~\eqref{eq:alpha0} this gives us the black solid line. The dashed gray lines are calculated from the error bars visible in Fig.~\ref{Fig:7}. For calculation of the black dashed line in (b) we use the same parameters, except that we assume that only atoms within the more narrow $\delta_D'=2\pi\times\qty{0.7}{MHz}$ have been inverted and the ground state atoms are removed [see Eq.~\eqref{eq:fze} for the distribution of the excited atoms]. The dash-dotted line in (b) is calculated under the same assumptions but without removing the ground state atoms [see Eq.~\eqref{eq:ground} for the distribution of the ground state atoms].}
\end{figure}

We now study absorption of the atomic beam when modifying the driving laser intensity. For this we measure $I_g/I_0$ on resonance, $\Delta_L=0$, and calculate the relative absorption with respect to an empty cavity. The experimental data are recorded using a heterodyne measurement, detailed in Sec.~\ref{sssec:2.1E}. The empty cavity signal is obtained by transferring the atoms to the storage states $\ket{m} = {}^3P_{0,2}$. We define the relative absorption as $(I_m - I_g)/I_m$.
We fix the oven temperature to $T=\qty{803}{K}$ and map out the absorption in Fig.~\ref{Fig:8}(a). We see that the amount of relative absorption decreases as saturation increases. In order to compare experimental data with theory, we use the fit parameters of Fig.~\ref{Fig:7} for $T=\qty{803}{K}$  $N\approx4.1\times10^5$, $\delta_D\approx2\pi\times\qty{2.38}{MHz}$, $\delta_0\approx2\pi\times\qty{1.74}{MHz}$. With these parameters, we solve Eq.~\eqref{eq:J} self-consistently with Eq.~\eqref{eq:alpha0} as a function of the intensity $I/I_{\mathrm{sat}}$. The result is visible as a black solid line and is in very good agreement with the experimental result over several orders of magnitude of the probe power.

Having demonstrated the transition to nonlinear behavior in the $\ket{g}$ atom-cavity system (bi-NMS), resulting in a $25\%$ absorption on resonance, we now consider an inverted-state atomic beam. This is done by introducing the storage, cleaning, and excitation stage to reach a $\ket{e}$ population that is spectrally separated from any remaining ground-state atoms. 
Cavity transmission measurements reveal the presence of optical gain for the inverted sample, as depicted in Fig.~\ref{Fig:8}(b). We define the relative gain as $(I_e - I_m)/I_m$. At low intensities, we observe a relative gain of $\sim10\%$, which decreases with increasing drive intensities. We use the same fitting parameters as above and solve Eq.~\eqref{eq:J} together with Eq.~\eqref{eq:alpha0} self-consistently. The resulting curve, represented by the solid black line in Fig.~\ref{Fig:8}(b), illustrates the achievable gain for perfect inversion of the atomic ensemble measured in Fig.~\ref{Fig:8}(a). In particular, this means that we assume $N_e=N$, $\delta_D^{(e)}=2\pi\times\qty{2.38}{MHz}$, and $\delta_0^{(e)}=2\pi\times\qty{3.53}{MHz}$. We see that this results in a gain that is a factor of two larger than the experimentally observed value. We expect this discrepancy to be caused by the limited efficiency with which we prepare atoms across all Doppler detunings. To incorporate this effect into the theory we assume that only a range of $\delta_D'=2\pi\times\qty{0.7}{MHz}$ around the center of the momentum distribution is inverted and all remaining ground state atoms are removed by the cleaning stage. The frequency range of $\delta_D'$ corresponds to the power broadening achieved in the storage stage discussed in Sec.~\ref{sssec:2.1B}. 
From this assumption, we can calculate the excited-state distribution function for the Doppler shift $u_z$ as
\begin{align}
    f_z^{(e)}(u_z)=&\frac{N_e}{\sqrt{2\pi(\delta_D^{(e)})^2}}e^{-\frac{(u_z-\delta_0^{(e)})^2}{2(\delta_D^{e})^2}}\nonumber\\
    =&\frac{N}{\sqrt{2\pi(\delta_D^{(g)})^2}}e^{-\frac{(u_z-\delta_0^{(g)})^2}{2\delta_D^2}}e^{-\frac{u_z^2}{2(\delta_D')^2}}.\label{eq:fze}
\end{align}
Using the above equality we find $N_e\approx0.1N$, $\delta_D^{(e)}\approx2\pi\times \qty{0.67}{MHz}$, and $\delta_0^{(e)}\approx2\pi\times\qty{0.28}{MHz}.$ We then use these parameters and set $N_g=0$ to calculate the expected gain. The result of this calculation is visible as a dashed line in Fig.~\ref{Fig:8}(b) and is still overestimating the gain observed in the experiment.

The reason for this overestimation is likely a combination of the need for a more precise description of the excitation (e.g., by incorporating also the velocity profile along the $x$-direction) and imperfect removal of ground-state atoms in the experiment. To incorporate the latter in our description we calculate the gain for a modified excited state distribution [see Eq.~\eqref{eq:fze}] but assume that none of the remaining ground-state atoms are removed by the cleaning stage such that
\begin{align}
    f_z^{(g)}(u_z)=\frac{N}{\sqrt{2\pi(\delta_D^{(g)})^2}}e^{-\frac{(u_z-\delta_0^{(g)})^2}{2(\delta_D)^2}}\left[1-e^{-\frac{u_z^2}{2(\delta_D')^2}}\right].\label{eq:ground}
\end{align}
This produces the expected gain visible as a dash-dotted line in Fig.~\ref{Fig:8}(b). The agreement with the experimental data indicates that residual ground state atoms, likely due to inefficiencies in the cleaning stage, may account for the observed gain reduction. However, we consider the complete inefficiency of the cleaning stage to be improbable and propose that the gain reduction has the additional cause of a broadening of the gain medium due to relative misalignments between and divergence of the storage laser, the atomic beam, and the cavity axes.  

\begin{figure}
	\includegraphics[width=1\linewidth]{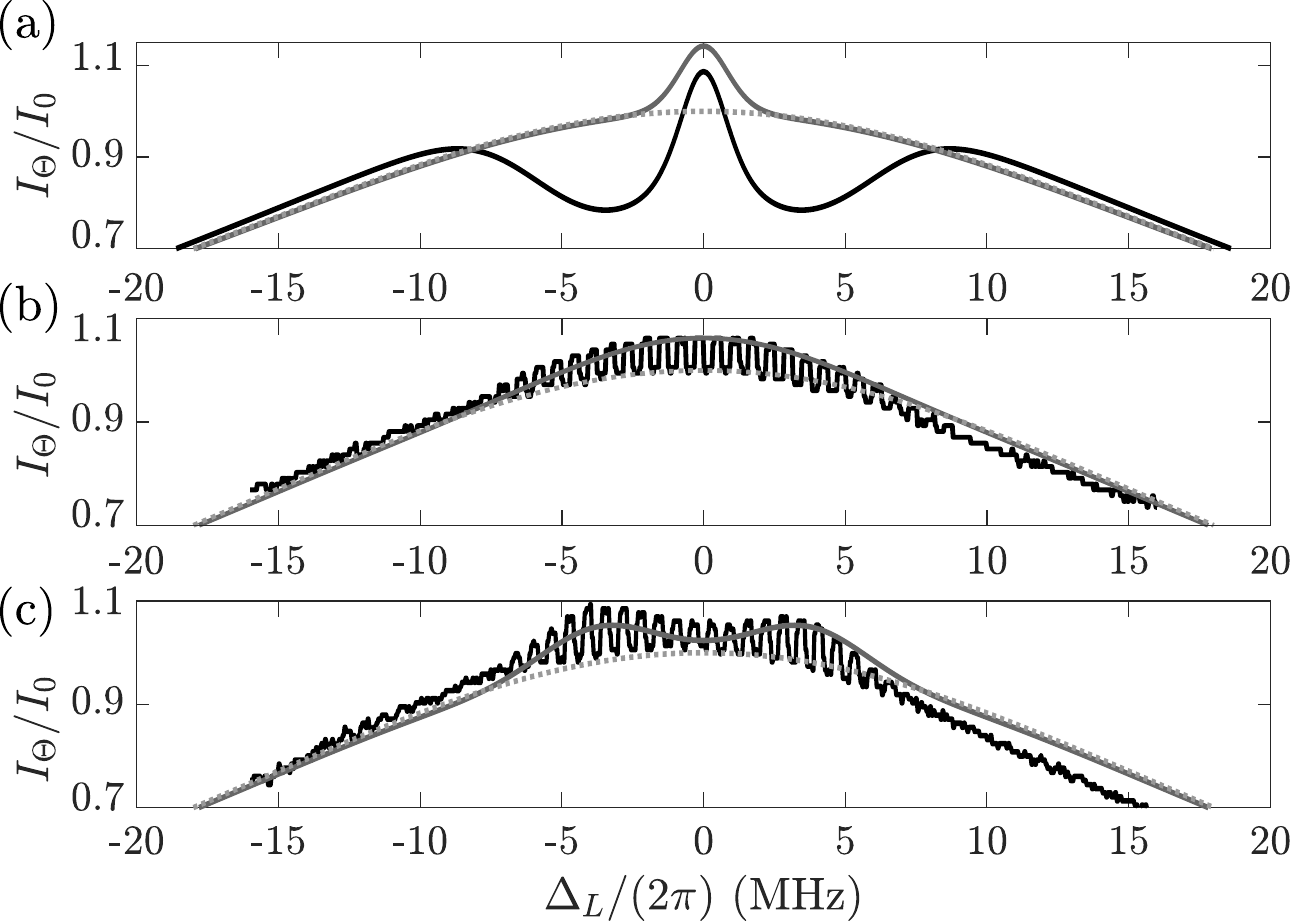}
	\caption{(a) Theoretical prediction of the gain spectrum, obtained using Eq.~\eqref{eq:ground} and Eq.~\eqref{eq:fze} in Eqs~\eqref{eq:deltaalpha} and \eqref{eq:Gamma}, shown as black solid line ($\Theta = e$) with $N=4.1\times10^5$, $\delta_D^{(g)}=2\pi\times \qty{2.38}{MHz}$, $\delta_0^{(g)}=2\pi\times\qty{3.53}{MHz}$, $T=\qty{803}{K}$, and $\delta_D'=2\pi\times\qty{0.7}{MHz}$. The gray solid line ($\Theta = e$) is shown for the same parameters but after removing the ground state atoms.  The experimentally observed gain spectra for (b) and (c) are shown as black solid lines. Their modulation is a result of alternating between atoms in $\ket{e}$ ($\Theta = e$) or $\ket{m}$ ($\Theta = m$). While the probe frequency is scanned over 30\,MHz in 100\,ms, optical pumping of atoms from $\ket{m}$ to $\ket{e}$ is toggled on or off with a period of 0.5\,ms. The experimental parameters are $T=\qty{803}{K}$ with an intracavity power corresponding to $I_0/I_{\mathrm{sat}}=1.1\times 10^4$. By comparing a Gaussian to the $\ket{m}$ signal we find that the asymmetric shape of the dataset is given by a non-zero cavity detuning of the order of 2$\pi \times $1.25\,MHz. The dark gray ($\Theta = e$) lines are theory calculations with $N_e = 1\times10^5$ and (b) $\delta_0^{(e)}=2\pi\times\qty{2.5}{MHz}$, $\delta_D^{(e)}=2\pi\times\qty{4}{MHz}$, and (c) $\delta_0^{(e)}=2\pi\times\qty{4}{MHz}$, $\delta_D^{(e)}=2\pi\times\qty{2}{MHz}$. In (a)-(c) we show the expected empty-cavity spectrum as a light gray dotted line.
 \label{Fig:9}}
\end{figure}

This claim is supported by studying the gain spectrum in Fig.~\ref{Fig:9}. In Fig.~\ref{Fig:9}(a), we present the predicted gain spectrum based on the theory used to calculate the gain shown in Fig.~\ref{Fig:8}(b). As a black solid line we show the case where atoms are excited within a frequency window of $\delta_D'=2\pi\times\qty{0.7}{MHz}$, producing a resonant gain peak, and where ground-state atoms are not removed, causing detuned absorption dips. By contrast, the gray line represents the same conditions with perfect removal of ground-state atoms. This results in the elimination of the absorption dips and predicts a higher gain at zero detuning $\Delta_L=0$ (compare with the dashed and dash-dotted line in Fig.~\ref{Fig:8}(b)). As a guide to the eye, we show the empty-cavity response as a light gray dotted line.

To compare our insights from the theoretical description with the measured data, we show in Fig.~\ref{Fig:9}(b) and (c) the measured transmission as black solid lines. Incoherent pumping to $\ket{e}$ from metastable states ${}^3P_{0,2}$ ($\ket{m}$) is modulated to compare, in real-time, an empty-cavity-like condition, $\Theta = m$ and a cavity with an inverted sample, $\Theta = e$.
In contrast to the theory, we do not find frequency ranges in which we observe large absorption from fast ground-state atoms, comparable to the gain.
However, we notice small effects of these atoms in both the $\Theta = e, m$ signals, where the transmission at high detunings ($\sim 10$~MHz) is reduced compared to the expected bare-cavity distribution.

Experimentally, we see an optical gain even for rather large detunings $\Delta_L>\delta_D'$, compared to the theoretical prediction visible in Fig.~\ref{Fig:9}(a). This is likely caused by a small divergence in the storage laser beam, broadening the range of velocities addressed. Furthermore, Figure~\ref{Fig:9}(b) shows the maximum gain at resonance $\Delta_L\approx0$, corresponding to the condition shown in Fig~\ref{Fig:8}(b), while Figure~\ref{Fig:9}(c) shows an off-resonant maximum $\Delta_L\approx2\pi\times\qty{4}{MHz}$. These features result from slight differences in the alignment of the preparation beams.
The latter figure is in its form reminiscent of our theoretical prediction of Fig.~\ref{Fig:4}, likely due to an offset of the stream of excited state atoms, $\delta_0^{(e)}\neq0$.
 
We show theoretical curves where we assume $N_e = 1\times10^5$, (b) $\delta_0^{(e)}=2\pi\times\qty{2.5}{MHz}$, $\delta_D^{(e)}=2\pi\times\qty{4}{MHz}$, and (c) $\delta_0^{(e)}=2\pi\times\qty{4}{MHz}$, $\delta_D^{(e)}=2\pi\times\qty{2}{MHz}$. 
In addition, we have removed all ground-state atoms in our simulation $N_g=0$. We point out that the choice of parameters is not unique and showcases how the gain spectrum can be influenced by variations in preparation beam conditions. 
From these observations and simulations, we can draw two conclusions. First, the small absorption effects indicate that a large portion of the ground-state atoms have been detuned, suggesting that the cleaning stage is partially effective. Second, the optical gain distribution in the cavity mode is subject to broadening and offsets, similar to the normal mode splitting, due to minor divergences and misalignments (less than 1 degree) respectively.

\section{\label{sec:5}Conclusions and future directions}

We have described the design and characterization of an apparatus suitable for performing continuous cavity-QED experiments with cold atoms in the short-memory-cavity regime. By preparing a continuous sample of ${}^1S_0$ $^{88}$Sr atoms that couple to the mode of a bad cavity, we have demonstrated the emergence of nonlinear behavior in the system on the ${}^1S_0-{}^3P_1$ transition. We see this in the form of the well-known normal mode splitting, but also observe a regime in which the system has a binary-type normal mode splitting. By inverting the atomic sample and driving the system with an external resonant seed, we observe an increase in transmission of up to $10\%$ compared to an empty cavity, indicating the presence of optical gain.
This signal constitutes the first steps towards realizing fully continuous superradiant lasing on the intercombination transition of strontium. 
In order to interpret the experimental results, we have developed a mean-field theory that comprehensively accounts for real-world effects such as transit time, Doppler broadening, Gaussian cavity mode, velocity offsets, and imperfect pumping. Moreover, we have investigated how various parameter regimes can be discerned through simple experimental measurements. Specifically, we have examined the scenarios of weakly or strongly driven atom-cavity systems as benchmarks for detecting the emergence of an atomic collective state and characterizing its properties. 
Our studies have also brought to light certain limitations of traditional cold-atom experimental apparatus designs. The observed discrepancy between the theoretically predicted and the experimentally observed optical gain values underscores the need to improve the atomic state preparation scheme. This invites a number of modifications to the experimental system, such as improvements to the transverse cooling and state-transfer stages.
Both the experimental and theoretical analysis indicate that we are only an order of magnitude away from achieving self-sustaining lasing conditions.

\begin{acknowledgments}

The authors wish to thank Haonan Liu, Premjith Thekkeppatt, Jinx Cooper, and Murray Holland for useful discussions. In addition, we thank the mechanical and electronics workshop of the University of Amsterdam for technical support and Camila Beli Silva for contributions to the experimental setup. This project has received funding from the European Union’s (EU) Horizon 2020 research and innovation program under Grant Agreement No. 820404 (iqClock project) and No. 860579 (MoSaiQ project). It further received funding from the Dutch National Growth Fund (NGF), as part of the Quantum Delta NL programme. S.B.J. acknowledges support from the Deutsche Forschungsgemeinschaft (DFG, German Research Foundation) through projects A4 and A5 in TRR-185 ``OSCAR''. S.A.S. has recieved funding from the Independent Research Fund Denmark for support under Project No. 0131-00023B and the European Union’s (EU) Horizon 2020 research and innovation program under the Marie Skłodowska Curie grant agreement No 101109698.

\textit{Author contributions ---} S.B. and F.F., assisted by S.A.S., S.Z., and M.T., developed the initial experimental design. F.F., S.Z., M.T., S.B., and S.A.S. constructed the apparatus. Initial debugging was done by F.F., S.Z. and S.B. Further key improvements to the apparatus were implemented by F.F., B.H., and S.A.S. Data taking, analysis and fundamental understanding were done by F.F., B.H. and S.A.S. S.B.J. developed the theoretical formalism, performed analytic calculations and carried out numerical simulations, supported by F.F. and S.A.S.
The manuscript was written by F.F., S.B.J., and S.A.S., with assistance from B.H. and F.S. The project was supervised by S.B., S.A.S and F.S. Funding was acquired by F.S.

\end{acknowledgments}

	\bibliography{bibliography}

 \newcommand{\noop}[1]{}
\begin{thebibliography}{35}%
\makeatletter
\providecommand \@ifxundefined [1]{%
 \@ifx{#1\undefined}
}%
\providecommand \@ifnum [1]{%
 \ifnum #1\expandafter \@firstoftwo
 \else \expandafter \@secondoftwo
 \fi
}%
\providecommand \@ifx [1]{%
 \ifx #1\expandafter \@firstoftwo
 \else \expandafter \@secondoftwo
 \fi
}%
\providecommand \natexlab [1]{#1}%
\providecommand \enquote  [1]{``#1''}%
\providecommand \bibnamefont  [1]{#1}%
\providecommand \bibfnamefont [1]{#1}%
\providecommand \citenamefont [1]{#1}%
\providecommand \href@noop [0]{\@secondoftwo}%
\providecommand \href [0]{\begingroup \@sanitize@url \@href}%
\providecommand \@href[1]{\@@startlink{#1}\@@href}%
\providecommand \@@href[1]{\endgroup#1\@@endlink}%
\providecommand \@sanitize@url [0]{\catcode `\\12\catcode `\$12\catcode `\&12\catcode `\#12\catcode `\^12\catcode `\_12\catcode `\%12\relax}%
\providecommand \@@startlink[1]{}%
\providecommand \@@endlink[0]{}%
\providecommand \url  [0]{\begingroup\@sanitize@url \@url }%
\providecommand \@url [1]{\endgroup\@href {#1}{\urlprefix }}%
\providecommand \urlprefix  [0]{URL }%
\providecommand \Eprint [0]{\href }%
\providecommand \doibase [0]{https://doi.org/}%
\providecommand \selectlanguage [0]{\@gobble}%
\providecommand \bibinfo  [0]{\@secondoftwo}%
\providecommand \bibfield  [0]{\@secondoftwo}%
\providecommand \translation [1]{[#1]}%
\providecommand \BibitemOpen [0]{}%
\providecommand \bibitemStop [0]{}%
\providecommand \bibitemNoStop [0]{.\EOS\space}%
\providecommand \EOS [0]{\spacefactor3000\relax}%
\providecommand \BibitemShut  [1]{\csname bibitem#1\endcsname}%
\let\auto@bib@innerbib\@empty
\bibitem [{\citenamefont {Dieckmann}\ \emph {et~al.}(1998)\citenamefont {Dieckmann}, \citenamefont {Spreeuw}, \citenamefont {Weidem\"uller},\ and\ \citenamefont {Walraven}}]{Dieckmann1998}%
  \BibitemOpen
  \bibfield  {author} {\bibinfo {author} {\bibfnamefont {K.}~\bibnamefont {Dieckmann}}, \bibinfo {author} {\bibfnamefont {R.~J.~C.}\ \bibnamefont {Spreeuw}}, \bibinfo {author} {\bibfnamefont {M.}~\bibnamefont {Weidem\"uller}},\ and\ \bibinfo {author} {\bibfnamefont {J.~T.~M.}\ \bibnamefont {Walraven}},\ }\bibfield  {title} {\bibinfo {title} {Two-dimensional magneto-optical trap as a source of slow atoms},\ }\href {https://doi.org/10.1103/PhysRevA.58.3891} {\bibfield  {journal} {\bibinfo  {journal} {Phys. Rev. A}\ }\textbf {\bibinfo {volume} {58}},\ \bibinfo {pages} {3891} (\bibinfo {year} {1998})}\BibitemShut {NoStop}%
\bibitem [{\citenamefont {Lahaye}\ \emph {et~al.}(2005)\citenamefont {Lahaye}, \citenamefont {Wang}, \citenamefont {Reinaudi}, \citenamefont {Rath}, \citenamefont {Dalibard},\ and\ \citenamefont {Gu\'ery-Odelin}}]{Lahaye2005}%
  \BibitemOpen
  \bibfield  {author} {\bibinfo {author} {\bibfnamefont {T.}~\bibnamefont {Lahaye}}, \bibinfo {author} {\bibfnamefont {Z.}~\bibnamefont {Wang}}, \bibinfo {author} {\bibfnamefont {G.}~\bibnamefont {Reinaudi}}, \bibinfo {author} {\bibfnamefont {S.~P.}\ \bibnamefont {Rath}}, \bibinfo {author} {\bibfnamefont {J.}~\bibnamefont {Dalibard}},\ and\ \bibinfo {author} {\bibfnamefont {D.}~\bibnamefont {Gu\'ery-Odelin}},\ }\bibfield  {title} {\bibinfo {title} {Evaporative cooling of a guided rubidium atomic beam},\ }\href {https://doi.org/10.1103/PhysRevA.72.033411} {\bibfield  {journal} {\bibinfo  {journal} {Phys. Rev. A}\ }\textbf {\bibinfo {volume} {72}},\ \bibinfo {pages} {033411} (\bibinfo {year} {2005})}\BibitemShut {NoStop}%
\bibitem [{\citenamefont {Chen}\ \emph {et~al.}(2022)\citenamefont {Chen}, \citenamefont {Gonz{\'a}lez~Escudero}, \citenamefont {Min{\'a}{\v{r}}}, \citenamefont {Pasquiou}, \citenamefont {Bennetts},\ and\ \citenamefont {Schreck}}]{Chen:2022}%
  \BibitemOpen
  \bibfield  {author} {\bibinfo {author} {\bibfnamefont {C.-C.}\ \bibnamefont {Chen}}, \bibinfo {author} {\bibfnamefont {R.}~\bibnamefont {Gonz{\'a}lez~Escudero}}, \bibinfo {author} {\bibfnamefont {J.}~\bibnamefont {Min{\'a}{\v{r}}}}, \bibinfo {author} {\bibfnamefont {B.}~\bibnamefont {Pasquiou}}, \bibinfo {author} {\bibfnamefont {S.}~\bibnamefont {Bennetts}},\ and\ \bibinfo {author} {\bibfnamefont {F.}~\bibnamefont {Schreck}},\ }\bibfield  {title} {\bibinfo {title} {Continuous {B}ose--{E}instein condensation},\ }\href {https://doi.org/10.1038/s41586-022-04731-z} {\bibfield  {journal} {\bibinfo  {journal} {Nature}\ }\textbf {\bibinfo {volume} {606}},\ \bibinfo {pages} {683} (\bibinfo {year} {2022})}\BibitemShut {NoStop}%
\bibitem [{\citenamefont {Huntington}\ \emph {et~al.}(2023)\citenamefont {Huntington}, \citenamefont {Glick}, \citenamefont {Borysow},\ and\ \citenamefont {Heinzen}}]{Huntington2023}%
  \BibitemOpen
  \bibfield  {author} {\bibinfo {author} {\bibfnamefont {W.}~\bibnamefont {Huntington}}, \bibinfo {author} {\bibfnamefont {J.}~\bibnamefont {Glick}}, \bibinfo {author} {\bibfnamefont {M.}~\bibnamefont {Borysow}},\ and\ \bibinfo {author} {\bibfnamefont {D.~J.}\ \bibnamefont {Heinzen}},\ }\bibfield  {title} {\bibinfo {title} {Intense continuous cold-atom source},\ }\href {https://doi.org/10.1103/PhysRevA.107.013302} {\bibfield  {journal} {\bibinfo  {journal} {Phys. Rev. A}\ }\textbf {\bibinfo {volume} {107}},\ \bibinfo {pages} {013302} (\bibinfo {year} {2023})}\BibitemShut {NoStop}%
\bibitem [{\citenamefont {Okaba}\ \emph {et~al.}(2024)\citenamefont {Okaba}, \citenamefont {Takeuchi}, \citenamefont {Tsuji},\ and\ \citenamefont {Katori}}]{Okaba:2024}%
  \BibitemOpen
  \bibfield  {author} {\bibinfo {author} {\bibfnamefont {S.}~\bibnamefont {Okaba}}, \bibinfo {author} {\bibfnamefont {R.}~\bibnamefont {Takeuchi}}, \bibinfo {author} {\bibfnamefont {S.}~\bibnamefont {Tsuji}},\ and\ \bibinfo {author} {\bibfnamefont {H.}~\bibnamefont {Katori}},\ }\bibfield  {title} {\bibinfo {title} {Continuous generation of an ultracold atomic beam using crossed moving optical lattices},\ }\href {https://doi.org/10.1103/PhysRevApplied.21.034006} {\bibfield  {journal} {\bibinfo  {journal} {Phys. Rev. Appl.}\ }\textbf {\bibinfo {volume} {21}},\ \bibinfo {pages} {034006} (\bibinfo {year} {2024})}\BibitemShut {NoStop}%
\bibitem [{\citenamefont {Sch{\"a}fer}\ \emph {et~al.}(2024)\citenamefont {Sch{\"a}fer}, \citenamefont {Niu}, \citenamefont {Cline}, \citenamefont {Young}, \citenamefont {Song}, \citenamefont {Ritsch},\ and\ \citenamefont {Thompson}}]{schafer:2024}%
  \BibitemOpen
  \bibfield  {author} {\bibinfo {author} {\bibfnamefont {V.}~\bibnamefont {Sch{\"a}fer}}, \bibinfo {author} {\bibfnamefont {Z.}~\bibnamefont {Niu}}, \bibinfo {author} {\bibfnamefont {J.}~\bibnamefont {Cline}}, \bibinfo {author} {\bibfnamefont {D.}~\bibnamefont {Young}}, \bibinfo {author} {\bibfnamefont {E.}~\bibnamefont {Song}}, \bibinfo {author} {\bibfnamefont {H.}~\bibnamefont {Ritsch}},\ and\ \bibinfo {author} {\bibfnamefont {J.}~\bibnamefont {Thompson}},\ }\bibfield  {title} {\bibinfo {title} {Continuous momentum state lasing and cavity frequency-pinning with laser-cooled strontium atoms},\ }\bibfield  {journal} {\bibinfo  {journal} {arXiv preprint arXiv:2405.20952}\ }\href {https://doi.org/10.48550/arXiv.2405.20952} {10.48550/arXiv.2405.20952} (\bibinfo {year} {2024})\BibitemShut {NoStop}%
\bibitem [{\citenamefont {Gyger}\ \emph {et~al.}(2024)\citenamefont {Gyger}, \citenamefont {Ammenwerth}, \citenamefont {Tao}, \citenamefont {Timme}, \citenamefont {Snigirev}, \citenamefont {Bloch},\ and\ \citenamefont {Zeiher}}]{Gyger2024}%
  \BibitemOpen
  \bibfield  {author} {\bibinfo {author} {\bibfnamefont {F.}~\bibnamefont {Gyger}}, \bibinfo {author} {\bibfnamefont {M.}~\bibnamefont {Ammenwerth}}, \bibinfo {author} {\bibfnamefont {R.}~\bibnamefont {Tao}}, \bibinfo {author} {\bibfnamefont {H.}~\bibnamefont {Timme}}, \bibinfo {author} {\bibfnamefont {S.}~\bibnamefont {Snigirev}}, \bibinfo {author} {\bibfnamefont {I.}~\bibnamefont {Bloch}},\ and\ \bibinfo {author} {\bibfnamefont {J.}~\bibnamefont {Zeiher}},\ }\bibfield  {title} {\bibinfo {title} {Continuous operation of large-scale atom arrays in optical lattices},\ }\href {https://doi.org/10.48550/arXiv.2402.04994} {\bibfield  {journal} {\bibinfo  {journal} {arXiv:2402.04994}\ } (\bibinfo {year} {2024})}\BibitemShut {NoStop}%
\bibitem [{\citenamefont {Norcia}\ \emph {et~al.}(2024)\citenamefont {Norcia}, \citenamefont {Kim}, \citenamefont {Cairncross}, \citenamefont {Stone}, \citenamefont {Ryou}, \citenamefont {Jaffe}, \citenamefont {Brown}, \citenamefont {Barnes}, \citenamefont {Battaglino}, \citenamefont {Bohdanowicz}, \citenamefont {Brown}, \citenamefont {Cassella}, \citenamefont {Chen}, \citenamefont {Coxe}, \citenamefont {Crow}, \citenamefont {Epstein}, \citenamefont {Griger}, \citenamefont {Halperin}, \citenamefont {Hummel}, \citenamefont {Jones}, \citenamefont {Kindem}, \citenamefont {King}, \citenamefont {Kotru}, \citenamefont {Lauigan}, \citenamefont {Li}, \citenamefont {Lu}, \citenamefont {Megidish}, \citenamefont {Marjanovic}, \citenamefont {McDonald}, \citenamefont {Mittiga}, \citenamefont {Muniz}, \citenamefont {Narayanaswami}, \citenamefont {Nishiguchi}, \citenamefont {Paule}, \citenamefont {Pawlak}, \citenamefont {Peng}, \citenamefont {Pudenz}, \citenamefont {Perez}, \citenamefont {Smull}, \citenamefont {Stack},
  \citenamefont {Urbanek}, \citenamefont {van~de Veerdonk}, \citenamefont {Vendeiro}, \citenamefont {Wadleigh}, \citenamefont {Wilkason}, \citenamefont {Wu}, \citenamefont {Xie}, \citenamefont {Zalys-Geller}, \citenamefont {Zhang},\ and\ \citenamefont {Bloom}}]{Norcia2024}%
  \BibitemOpen
  \bibfield  {author} {\bibinfo {author} {\bibfnamefont {M.~A.}\ \bibnamefont {Norcia}}, \bibinfo {author} {\bibfnamefont {H.}~\bibnamefont {Kim}}, \bibinfo {author} {\bibfnamefont {W.~B.}\ \bibnamefont {Cairncross}}, \bibinfo {author} {\bibfnamefont {M.}~\bibnamefont {Stone}}, \bibinfo {author} {\bibfnamefont {A.}~\bibnamefont {Ryou}}, \bibinfo {author} {\bibfnamefont {M.}~\bibnamefont {Jaffe}}, \bibinfo {author} {\bibfnamefont {M.~O.}\ \bibnamefont {Brown}}, \bibinfo {author} {\bibfnamefont {K.}~\bibnamefont {Barnes}}, \bibinfo {author} {\bibfnamefont {P.}~\bibnamefont {Battaglino}}, \bibinfo {author} {\bibfnamefont {T.~C.}\ \bibnamefont {Bohdanowicz}}, \bibinfo {author} {\bibfnamefont {A.}~\bibnamefont {Brown}}, \bibinfo {author} {\bibfnamefont {K.}~\bibnamefont {Cassella}}, \bibinfo {author} {\bibfnamefont {C.~A.}\ \bibnamefont {Chen}}, \bibinfo {author} {\bibfnamefont {R.}~\bibnamefont {Coxe}}, \bibinfo {author} {\bibfnamefont {D.}~\bibnamefont {Crow}}, \bibinfo {author} {\bibfnamefont {J.}~\bibnamefont
  {Epstein}}, \bibinfo {author} {\bibfnamefont {C.}~\bibnamefont {Griger}}, \bibinfo {author} {\bibfnamefont {E.}~\bibnamefont {Halperin}}, \bibinfo {author} {\bibfnamefont {F.}~\bibnamefont {Hummel}}, \bibinfo {author} {\bibfnamefont {A.~M.~W.}\ \bibnamefont {Jones}}, \bibinfo {author} {\bibfnamefont {J.~M.}\ \bibnamefont {Kindem}}, \bibinfo {author} {\bibfnamefont {J.}~\bibnamefont {King}}, \bibinfo {author} {\bibfnamefont {K.}~\bibnamefont {Kotru}}, \bibinfo {author} {\bibfnamefont {J.}~\bibnamefont {Lauigan}}, \bibinfo {author} {\bibfnamefont {M.}~\bibnamefont {Li}}, \bibinfo {author} {\bibfnamefont {M.}~\bibnamefont {Lu}}, \bibinfo {author} {\bibfnamefont {E.}~\bibnamefont {Megidish}}, \bibinfo {author} {\bibfnamefont {J.}~\bibnamefont {Marjanovic}}, \bibinfo {author} {\bibfnamefont {M.}~\bibnamefont {McDonald}}, \bibinfo {author} {\bibfnamefont {T.}~\bibnamefont {Mittiga}}, \bibinfo {author} {\bibfnamefont {J.~A.}\ \bibnamefont {Muniz}}, \bibinfo {author} {\bibfnamefont {S.}~\bibnamefont
  {Narayanaswami}}, \bibinfo {author} {\bibfnamefont {C.}~\bibnamefont {Nishiguchi}}, \bibinfo {author} {\bibfnamefont {T.}~\bibnamefont {Paule}}, \bibinfo {author} {\bibfnamefont {K.~A.}\ \bibnamefont {Pawlak}}, \bibinfo {author} {\bibfnamefont {L.~S.}\ \bibnamefont {Peng}}, \bibinfo {author} {\bibfnamefont {K.~L.}\ \bibnamefont {Pudenz}}, \bibinfo {author} {\bibfnamefont {D.~R.}\ \bibnamefont {Perez}}, \bibinfo {author} {\bibfnamefont {A.}~\bibnamefont {Smull}}, \bibinfo {author} {\bibfnamefont {D.}~\bibnamefont {Stack}}, \bibinfo {author} {\bibfnamefont {M.}~\bibnamefont {Urbanek}}, \bibinfo {author} {\bibfnamefont {R.~J.~M.}\ \bibnamefont {van~de Veerdonk}}, \bibinfo {author} {\bibfnamefont {Z.}~\bibnamefont {Vendeiro}}, \bibinfo {author} {\bibfnamefont {L.}~\bibnamefont {Wadleigh}}, \bibinfo {author} {\bibfnamefont {T.}~\bibnamefont {Wilkason}}, \bibinfo {author} {\bibfnamefont {T.~Y.}\ \bibnamefont {Wu}}, \bibinfo {author} {\bibfnamefont {X.}~\bibnamefont {Xie}}, \bibinfo {author} {\bibfnamefont
  {E.}~\bibnamefont {Zalys-Geller}}, \bibinfo {author} {\bibfnamefont {X.}~\bibnamefont {Zhang}},\ and\ \bibinfo {author} {\bibfnamefont {B.~J.}\ \bibnamefont {Bloom}},\ }\bibfield  {title} {\bibinfo {title} {Iterative assembly of $^{171}${Y}b atom arrays in cavity-enhanced optical lattices},\ }\href {https://doi.org/10.48550/arXiv.2401.16177} {\bibfield  {journal} {\bibinfo  {journal} {arXiv:2401.16177}\ } (\bibinfo {year} {2024})}\BibitemShut {NoStop}%
\bibitem [{\citenamefont {Acín}\ \emph {et~al.}(2018)\citenamefont {Acín}, \citenamefont {Bloch}, \citenamefont {Buhrman}, \citenamefont {Calarco}, \citenamefont {Eichler}, \citenamefont {Eisert}, \citenamefont {Esteve}, \citenamefont {Gisin}, \citenamefont {Glaser}, \citenamefont {Jelezko}, \citenamefont {Kuhr}, \citenamefont {Lewenstein}, \citenamefont {Riedel}, \citenamefont {Schmidt}, \citenamefont {Thew}, \citenamefont {Wallraff}, \citenamefont {Walmsley},\ and\ \citenamefont {Wilhelm}}]{Acin:2018}%
  \BibitemOpen
  \bibfield  {author} {\bibinfo {author} {\bibfnamefont {A.}~\bibnamefont {Acín}}, \bibinfo {author} {\bibfnamefont {I.}~\bibnamefont {Bloch}}, \bibinfo {author} {\bibfnamefont {H.}~\bibnamefont {Buhrman}}, \bibinfo {author} {\bibfnamefont {T.}~\bibnamefont {Calarco}}, \bibinfo {author} {\bibfnamefont {C.}~\bibnamefont {Eichler}}, \bibinfo {author} {\bibfnamefont {J.}~\bibnamefont {Eisert}}, \bibinfo {author} {\bibfnamefont {D.}~\bibnamefont {Esteve}}, \bibinfo {author} {\bibfnamefont {N.}~\bibnamefont {Gisin}}, \bibinfo {author} {\bibfnamefont {S.~J.}\ \bibnamefont {Glaser}}, \bibinfo {author} {\bibfnamefont {F.}~\bibnamefont {Jelezko}}, \bibinfo {author} {\bibfnamefont {S.}~\bibnamefont {Kuhr}}, \bibinfo {author} {\bibfnamefont {M.}~\bibnamefont {Lewenstein}}, \bibinfo {author} {\bibfnamefont {M.~F.}\ \bibnamefont {Riedel}}, \bibinfo {author} {\bibfnamefont {P.~O.}\ \bibnamefont {Schmidt}}, \bibinfo {author} {\bibfnamefont {R.}~\bibnamefont {Thew}}, \bibinfo {author} {\bibfnamefont {A.}~\bibnamefont
  {Wallraff}}, \bibinfo {author} {\bibfnamefont {I.}~\bibnamefont {Walmsley}},\ and\ \bibinfo {author} {\bibfnamefont {F.~K.}\ \bibnamefont {Wilhelm}},\ }\bibfield  {title} {\bibinfo {title} {The quantum technologies roadmap: a {E}uropean community view},\ }\href {https://doi.org/10.1088/1367-2630/aad1ea} {\bibfield  {journal} {\bibinfo  {journal} {New Journal of Physics}\ }\textbf {\bibinfo {volume} {20}},\ \bibinfo {pages} {080201} (\bibinfo {year} {2018})}\BibitemShut {NoStop}%
\bibitem [{\citenamefont {Xu}\ \emph {et~al.}(2024)\citenamefont {Xu}, \citenamefont {Bonilla~Ataides}, \citenamefont {Pattison}, \citenamefont {Raveendran}, \citenamefont {Bluvstein}, \citenamefont {Wurtz}, \citenamefont {Vasi{\'{c}}}, \citenamefont {Lukin}, \citenamefont {Jiang},\ and\ \citenamefont {Zhou}}]{Xu:2024}%
  \BibitemOpen
  \bibfield  {author} {\bibinfo {author} {\bibfnamefont {Q.}~\bibnamefont {Xu}}, \bibinfo {author} {\bibfnamefont {J.~P.}\ \bibnamefont {Bonilla~Ataides}}, \bibinfo {author} {\bibfnamefont {C.~A.}\ \bibnamefont {Pattison}}, \bibinfo {author} {\bibfnamefont {N.}~\bibnamefont {Raveendran}}, \bibinfo {author} {\bibfnamefont {D.}~\bibnamefont {Bluvstein}}, \bibinfo {author} {\bibfnamefont {J.}~\bibnamefont {Wurtz}}, \bibinfo {author} {\bibfnamefont {B.}~\bibnamefont {Vasi{\'{c}}}}, \bibinfo {author} {\bibfnamefont {M.~D.}\ \bibnamefont {Lukin}}, \bibinfo {author} {\bibfnamefont {L.}~\bibnamefont {Jiang}},\ and\ \bibinfo {author} {\bibfnamefont {H.}~\bibnamefont {Zhou}},\ }\bibfield  {title} {\bibinfo {title} {Constant-overhead fault-tolerant quantum computation with reconfigurable atom arrays},\ }\bibfield  {journal} {\bibinfo  {journal} {Nature Physics}\ }\href {https://doi.org/10.1038/s41567-024-02479-z} {10.1038/s41567-024-02479-z} (\bibinfo {year} {2024})\BibitemShut {NoStop}%
\bibitem [{\citenamefont {Olson}\ \emph {et~al.}(2019)\citenamefont {Olson}, \citenamefont {Fox}, \citenamefont {Fortier}, \citenamefont {Sheerin}, \citenamefont {Brown}, \citenamefont {Leopardi}, \citenamefont {Stoner}, \citenamefont {Oates},\ and\ \citenamefont {Ludlow}}]{Olson2019}%
  \BibitemOpen
  \bibfield  {author} {\bibinfo {author} {\bibfnamefont {J.}~\bibnamefont {Olson}}, \bibinfo {author} {\bibfnamefont {R.~W.}\ \bibnamefont {Fox}}, \bibinfo {author} {\bibfnamefont {T.~M.}\ \bibnamefont {Fortier}}, \bibinfo {author} {\bibfnamefont {T.~F.}\ \bibnamefont {Sheerin}}, \bibinfo {author} {\bibfnamefont {R.~C.}\ \bibnamefont {Brown}}, \bibinfo {author} {\bibfnamefont {H.}~\bibnamefont {Leopardi}}, \bibinfo {author} {\bibfnamefont {R.~E.}\ \bibnamefont {Stoner}}, \bibinfo {author} {\bibfnamefont {C.~W.}\ \bibnamefont {Oates}},\ and\ \bibinfo {author} {\bibfnamefont {A.~D.}\ \bibnamefont {Ludlow}},\ }\bibfield  {title} {\bibinfo {title} {Ramsey-{B}ord\'e matter-wave interferometry for laser frequency stabilization at ${10}^{\ensuremath{-}16}$ frequency instability and below},\ }\href {https://doi.org/10.1103/PhysRevLett.123.073202} {\bibfield  {journal} {\bibinfo  {journal} {Phys. Rev. Lett.}\ }\textbf {\bibinfo {volume} {123}},\ \bibinfo {pages} {073202} (\bibinfo {year} {2019})}\BibitemShut {NoStop}%
\bibitem [{\citenamefont {Katori}(2021)}]{Katori2021}%
  \BibitemOpen
  \bibfield  {author} {\bibinfo {author} {\bibfnamefont {H.}~\bibnamefont {Katori}},\ }\bibfield  {title} {\bibinfo {title} {Longitudinal {R}amsey spectroscopy of atoms for continuous operation of optical clocks},\ }\href {https://doi.org/10.35848/1882-0786/ac0e16} {\bibfield  {journal} {\bibinfo  {journal} {Applied Physics Express}\ }\textbf {\bibinfo {volume} {14}},\ \bibinfo {pages} {072006} (\bibinfo {year} {2021})}\BibitemShut {NoStop}%
\bibitem [{\citenamefont {Takeuchi}\ \emph {et~al.}(2023)\citenamefont {Takeuchi}, \citenamefont {Chiba}, \citenamefont {Okaba}, \citenamefont {Takamoto}, \citenamefont {Tsuji},\ and\ \citenamefont {Katori}}]{Takeuchi2023}%
  \BibitemOpen
  \bibfield  {author} {\bibinfo {author} {\bibfnamefont {R.}~\bibnamefont {Takeuchi}}, \bibinfo {author} {\bibfnamefont {H.}~\bibnamefont {Chiba}}, \bibinfo {author} {\bibfnamefont {S.}~\bibnamefont {Okaba}}, \bibinfo {author} {\bibfnamefont {M.}~\bibnamefont {Takamoto}}, \bibinfo {author} {\bibfnamefont {S.}~\bibnamefont {Tsuji}},\ and\ \bibinfo {author} {\bibfnamefont {H.}~\bibnamefont {Katori}},\ }\bibfield  {title} {\bibinfo {title} {Continuous outcoupling of ultracold strontium atoms combining three different traps},\ }\href {https://doi.org/10.35848/1882-0786/accb3c} {\bibfield  {journal} {\bibinfo  {journal} {Applied Physics Express}\ }\textbf {\bibinfo {volume} {16}},\ \bibinfo {pages} {042003} (\bibinfo {year} {2023})}\BibitemShut {NoStop}%
\bibitem [{\citenamefont {Carmichael}(1988)}]{Carmichael88}%
  \BibitemOpen
  \bibfield  {author} {\bibinfo {author} {\bibfnamefont {H.~J.}\ \bibnamefont {Carmichael}},\ }\bibfield  {title} {\bibinfo {title} {Antibunched light source using cavity-enhanced emission},\ }in\ \href {https://doi.org/10.1364/OAM.1988.MM5} {\emph {\bibinfo {booktitle} {Annual Meeting Optical Society of America}}}\ (\bibinfo  {publisher} {Optica Publishing Group},\ \bibinfo {year} {1988})\ p.\ \bibinfo {pages} {MM5}\BibitemShut {NoStop}%
\bibitem [{\citenamefont {Liu}\ \emph {et~al.}(2020)\citenamefont {Liu}, \citenamefont {J\"ager}, \citenamefont {Yu}, \citenamefont {Touzard}, \citenamefont {Shankar}, \citenamefont {Holland},\ and\ \citenamefont {Nicholson}}]{Liu:2020}%
  \BibitemOpen
  \bibfield  {author} {\bibinfo {author} {\bibfnamefont {H.}~\bibnamefont {Liu}}, \bibinfo {author} {\bibfnamefont {S.~B.}\ \bibnamefont {J\"ager}}, \bibinfo {author} {\bibfnamefont {X.}~\bibnamefont {Yu}}, \bibinfo {author} {\bibfnamefont {S.}~\bibnamefont {Touzard}}, \bibinfo {author} {\bibfnamefont {A.}~\bibnamefont {Shankar}}, \bibinfo {author} {\bibfnamefont {M.~J.}\ \bibnamefont {Holland}},\ and\ \bibinfo {author} {\bibfnamefont {T.~L.}\ \bibnamefont {Nicholson}},\ }\bibfield  {title} {\bibinfo {title} {Rugged m{H}z-linewidth superradiant laser driven by a hot atomic beam},\ }\href {https://doi.org/10.1103/PhysRevLett.125.253602} {\bibfield  {journal} {\bibinfo  {journal} {Phys. Rev. Lett.}\ }\textbf {\bibinfo {volume} {125}},\ \bibinfo {pages} {253602} (\bibinfo {year} {2020})}\BibitemShut {NoStop}%
\bibitem [{\citenamefont {Tang}\ \emph {et~al.}(2022)\citenamefont {Tang}, \citenamefont {Sch\"affer},\ and\ \citenamefont {M\"uller}}]{Tang:2022}%
  \BibitemOpen
  \bibfield  {author} {\bibinfo {author} {\bibfnamefont {M.}~\bibnamefont {Tang}}, \bibinfo {author} {\bibfnamefont {S.~A.}\ \bibnamefont {Sch\"affer}},\ and\ \bibinfo {author} {\bibfnamefont {J.~H.}\ \bibnamefont {M\"uller}},\ }\bibfield  {title} {\bibinfo {title} {Prospects of a superradiant laser based on a thermal or guided beam of $^{88}\mathrm{Sr}$},\ }\href {https://doi.org/10.1103/PhysRevA.106.063704} {\bibfield  {journal} {\bibinfo  {journal} {Phys. Rev. A}\ }\textbf {\bibinfo {volume} {106}},\ \bibinfo {pages} {063704} (\bibinfo {year} {2022})}\BibitemShut {NoStop}%
\bibitem [{\citenamefont {Numata}\ \emph {et~al.}(2004)\citenamefont {Numata}, \citenamefont {Kemery},\ and\ \citenamefont {Camp}}]{Numata2004}%
  \BibitemOpen
  \bibfield  {author} {\bibinfo {author} {\bibfnamefont {K.}~\bibnamefont {Numata}}, \bibinfo {author} {\bibfnamefont {A.}~\bibnamefont {Kemery}},\ and\ \bibinfo {author} {\bibfnamefont {J.}~\bibnamefont {Camp}},\ }\bibfield  {title} {\bibinfo {title} {Thermal-noise limit in the frequency stabilization of lasers with rigid cavities},\ }\href {https://doi.org/10.1103/PhysRevLett.93.250602} {\bibfield  {journal} {\bibinfo  {journal} {Phys. Rev. Lett.}\ }\textbf {\bibinfo {volume} {93}},\ \bibinfo {pages} {250602} (\bibinfo {year} {2004})}\BibitemShut {NoStop}%
\bibitem [{\citenamefont {Meiser}\ \emph {et~al.}(2009)\citenamefont {Meiser}, \citenamefont {Ye}, \citenamefont {Carlson},\ and\ \citenamefont {Holland}}]{Meiser:2009}%
  \BibitemOpen
  \bibfield  {author} {\bibinfo {author} {\bibfnamefont {D.}~\bibnamefont {Meiser}}, \bibinfo {author} {\bibfnamefont {J.}~\bibnamefont {Ye}}, \bibinfo {author} {\bibfnamefont {D.~R.}\ \bibnamefont {Carlson}},\ and\ \bibinfo {author} {\bibfnamefont {M.~J.}\ \bibnamefont {Holland}},\ }\bibfield  {title} {\bibinfo {title} {Prospects for a millihertz-linewidth laser},\ }\href {https://doi.org/10.1103/PhysRevLett.102.163601} {\bibfield  {journal} {\bibinfo  {journal} {Phys. Rev. Lett.}\ }\textbf {\bibinfo {volume} {102}},\ \bibinfo {pages} {163601} (\bibinfo {year} {2009})}\BibitemShut {NoStop}%
\bibitem [{\citenamefont {Dick}(1987)}]{Dick1987}%
  \BibitemOpen
  \bibfield  {author} {\bibinfo {author} {\bibfnamefont {G.~J.}\ \bibnamefont {Dick}},\ }\bibfield  {title} {\bibinfo {title} {Local oscillator induced instabilities in trapped ion frequency standards},\ }\href {;https://tycho.usno.navy.mil/ptti/1987papers/Vol%2019_13.pdf.} {\bibfield  {journal} {\bibinfo  {journal} {Proceedings of the 34th Annual Precise Time and Time Interval Systems and Applications Meeting (ION, 1987)}\ ,\ \bibinfo {pages} {133}} (\bibinfo {year} {1987})}\BibitemShut {NoStop}%
\bibitem [{\citenamefont {Westergaard}\ \emph {et~al.}(2010)\citenamefont {Westergaard}, \citenamefont {Lodewyck},\ and\ \citenamefont {Lemonde}}]{Westergaard2010}%
  \BibitemOpen
  \bibfield  {author} {\bibinfo {author} {\bibfnamefont {P.~G.}\ \bibnamefont {Westergaard}}, \bibinfo {author} {\bibfnamefont {J.}~\bibnamefont {Lodewyck}},\ and\ \bibinfo {author} {\bibfnamefont {P.}~\bibnamefont {Lemonde}},\ }\bibfield  {title} {\bibinfo {title} {Minimizing the {D}ick effect in an optical lattice clock},\ }\href {https://doi.org/10.1109/TUFFC.2010.1457} {\bibfield  {journal} {\bibinfo  {journal} {IEEE Transactions on Ultrasonics, Ferroelectrics, and Frequency Control}\ }\textbf {\bibinfo {volume} {57}},\ \bibinfo {pages} {623} (\bibinfo {year} {2010})}\BibitemShut {NoStop}%
\bibitem [{\citenamefont {Kim}\ \emph {et~al.}(2018)\citenamefont {Kim}, \citenamefont {Yang}, \citenamefont {hoon Oh},\ and\ \citenamefont {An}}]{Kim2017}%
  \BibitemOpen
  \bibfield  {author} {\bibinfo {author} {\bibfnamefont {J.}~\bibnamefont {Kim}}, \bibinfo {author} {\bibfnamefont {D.}~\bibnamefont {Yang}}, \bibinfo {author} {\bibfnamefont {S.}~\bibnamefont {hoon Oh}},\ and\ \bibinfo {author} {\bibfnamefont {K.}~\bibnamefont {An}},\ }\bibfield  {title} {\bibinfo {title} {Coherent single-atom superradiance},\ }\href {https://doi.org/10.1126/science.aar2179} {\bibfield  {journal} {\bibinfo  {journal} {Science}\ }\textbf {\bibinfo {volume} {359}},\ \bibinfo {pages} {662} (\bibinfo {year} {2018})}\BibitemShut {NoStop}%
\bibitem [{\citenamefont {Kristensen}\ \emph {et~al.}(2023)\citenamefont {Kristensen}, \citenamefont {Bohr}, \citenamefont {Robinson-Tait}, \citenamefont {Zelevinsky}, \citenamefont {Thomsen},\ and\ \citenamefont {M\"uller}}]{Kristensen:2023}%
  \BibitemOpen
  \bibfield  {author} {\bibinfo {author} {\bibfnamefont {S.~L.}\ \bibnamefont {Kristensen}}, \bibinfo {author} {\bibfnamefont {E.}~\bibnamefont {Bohr}}, \bibinfo {author} {\bibfnamefont {J.}~\bibnamefont {Robinson-Tait}}, \bibinfo {author} {\bibfnamefont {T.}~\bibnamefont {Zelevinsky}}, \bibinfo {author} {\bibfnamefont {J.~W.}\ \bibnamefont {Thomsen}},\ and\ \bibinfo {author} {\bibfnamefont {J.~H.}\ \bibnamefont {M\"uller}},\ }\bibfield  {title} {\bibinfo {title} {Subnatural linewidth superradiant lasing with cold $^{88}\mathrm{Sr}$ atoms},\ }\href {https://doi.org/10.1103/PhysRevLett.130.223402} {\bibfield  {journal} {\bibinfo  {journal} {Phys. Rev. Lett.}\ }\textbf {\bibinfo {volume} {130}},\ \bibinfo {pages} {223402} (\bibinfo {year} {2023})}\BibitemShut {NoStop}%
\bibitem [{\citenamefont {Norcia}\ \emph {et~al.}(2018)\citenamefont {Norcia}, \citenamefont {Lewis-Swan}, \citenamefont {Cline}, \citenamefont {Zhu}, \citenamefont {Rey},\ and\ \citenamefont {Thompson}}]{Norcia:2018:2}%
  \BibitemOpen
  \bibfield  {author} {\bibinfo {author} {\bibfnamefont {M.~A.}\ \bibnamefont {Norcia}}, \bibinfo {author} {\bibfnamefont {R.~J.}\ \bibnamefont {Lewis-Swan}}, \bibinfo {author} {\bibfnamefont {J.~R.~K.}\ \bibnamefont {Cline}}, \bibinfo {author} {\bibfnamefont {B.}~\bibnamefont {Zhu}}, \bibinfo {author} {\bibfnamefont {A.~M.}\ \bibnamefont {Rey}},\ and\ \bibinfo {author} {\bibfnamefont {J.~K.}\ \bibnamefont {Thompson}},\ }\bibfield  {title} {\bibinfo {title} {Cavity-mediated collective spin-exchange interactions in a strontium superradiant laser},\ }\href {https://doi.org/10.1126/science.aar3102} {\bibfield  {journal} {\bibinfo  {journal} {Science}\ }\textbf {\bibinfo {volume} {361}},\ \bibinfo {pages} {259} (\bibinfo {year} {2018})}\BibitemShut {NoStop}%
\bibitem [{\citenamefont {Cline}\ \emph {et~al.}(2022)\citenamefont {Cline}, \citenamefont {Sch{\"a}fer}, \citenamefont {Niu}, \citenamefont {Young}, \citenamefont {Yoon},\ and\ \citenamefont {Thompson}}]{cline:2022}%
  \BibitemOpen
  \bibfield  {author} {\bibinfo {author} {\bibfnamefont {J.~R.}\ \bibnamefont {Cline}}, \bibinfo {author} {\bibfnamefont {V.~M.}\ \bibnamefont {Sch{\"a}fer}}, \bibinfo {author} {\bibfnamefont {Z.}~\bibnamefont {Niu}}, \bibinfo {author} {\bibfnamefont {D.~J.}\ \bibnamefont {Young}}, \bibinfo {author} {\bibfnamefont {T.~H.}\ \bibnamefont {Yoon}},\ and\ \bibinfo {author} {\bibfnamefont {J.~K.}\ \bibnamefont {Thompson}},\ }\bibfield  {title} {\bibinfo {title} {Continuous collective strong coupling between atoms and a high finesse optical cavity},\ }\bibfield  {journal} {\bibinfo  {journal} {arXiv preprint arXiv:2211.00158}\ }\href {https://doi.org/10.48550/arXiv.2211.00158} {10.48550/arXiv.2211.00158} (\bibinfo {year} {2022})\BibitemShut {NoStop}%
\bibitem [{\citenamefont {Gea-Banacloche}\ \emph {et~al.}(2008)\citenamefont {Gea-Banacloche}, \citenamefont {Wu},\ and\ \citenamefont {Xiao}}]{gea2008}%
  \BibitemOpen
  \bibfield  {author} {\bibinfo {author} {\bibfnamefont {J.}~\bibnamefont {Gea-Banacloche}}, \bibinfo {author} {\bibfnamefont {H.}~\bibnamefont {Wu}},\ and\ \bibinfo {author} {\bibfnamefont {M.}~\bibnamefont {Xiao}},\ }\bibfield  {title} {\bibinfo {title} {Transmission spectrum of {D}oppler-broadened two-level atoms in a cavity in the strong-coupling regime},\ }\href {https://doi.org/10.1103/PhysRevA.78.023828} {\bibfield  {journal} {\bibinfo  {journal} {Phys. Rev. A}\ }\textbf {\bibinfo {volume} {78}},\ \bibinfo {pages} {023828} (\bibinfo {year} {2008})}\BibitemShut {NoStop}%
\bibitem [{\citenamefont {Zhang}\ \emph {et~al.}(2023)\citenamefont {Zhang}, \citenamefont {Shi}, \citenamefont {Miao},\ and\ \citenamefont {Chen}}]{zhang2023}%
  \BibitemOpen
  \bibfield  {author} {\bibinfo {author} {\bibfnamefont {J.}~\bibnamefont {Zhang}}, \bibinfo {author} {\bibfnamefont {T.}~\bibnamefont {Shi}}, \bibinfo {author} {\bibfnamefont {J.}~\bibnamefont {Miao}},\ and\ \bibinfo {author} {\bibfnamefont {J.}~\bibnamefont {Chen}},\ }\bibfield  {title} {\bibinfo {title} {The development of active optical clock},\ }\href@noop {} {\bibfield  {journal} {\bibinfo  {journal} {AAPPS Bulletin}\ }\textbf {\bibinfo {volume} {33}},\ \bibinfo {pages} {10} (\bibinfo {year} {2023})}\BibitemShut {NoStop}%
\bibitem [{\citenamefont {Nicholson}\ \emph {et~al.}(2015)\citenamefont {Nicholson}, \citenamefont {Campbell}, \citenamefont {Hutson}, \citenamefont {Marti}, \citenamefont {Bloom}, \citenamefont {McNally}, \citenamefont {Zhang}, \citenamefont {Barrett}, \citenamefont {Safronova}, \citenamefont {Strouse} \emph {et~al.}}]{nicholson:2015}%
  \BibitemOpen
  \bibfield  {author} {\bibinfo {author} {\bibfnamefont {T.~L.}\ \bibnamefont {Nicholson}}, \bibinfo {author} {\bibfnamefont {S.}~\bibnamefont {Campbell}}, \bibinfo {author} {\bibfnamefont {R.}~\bibnamefont {Hutson}}, \bibinfo {author} {\bibfnamefont {G.~E.}\ \bibnamefont {Marti}}, \bibinfo {author} {\bibfnamefont {B.}~\bibnamefont {Bloom}}, \bibinfo {author} {\bibfnamefont {R.~L.}\ \bibnamefont {McNally}}, \bibinfo {author} {\bibfnamefont {W.}~\bibnamefont {Zhang}}, \bibinfo {author} {\bibfnamefont {M.}~\bibnamefont {Barrett}}, \bibinfo {author} {\bibfnamefont {M.~S.}\ \bibnamefont {Safronova}}, \bibinfo {author} {\bibfnamefont {G.}~\bibnamefont {Strouse}}, \emph {et~al.},\ }\bibfield  {title} {\bibinfo {title} {Systematic evaluation of an atomic clock at 2$\times 10^{-18}$ total uncertainty},\ }\href {https://doi.org/10.1038/ncomms7896} {\bibfield  {journal} {\bibinfo  {journal} {Nature communications}\ }\textbf {\bibinfo {volume} {6}},\ \bibinfo {pages} {1} (\bibinfo {year} {2015})}\BibitemShut {NoStop}%
\bibitem [{\citenamefont {Arne~Wickenbrock}\ and\ \citenamefont {Renzoni}(2011)}]{Wickenbrock2011}%
  \BibitemOpen
  \bibfield  {author} {\bibinfo {author} {\bibfnamefont {P.~P.}\ \bibnamefont {Arne~Wickenbrock}}\ and\ \bibinfo {author} {\bibfnamefont {F.}~\bibnamefont {Renzoni}},\ }\bibfield  {title} {\bibinfo {title} {Collective strong coupling in a lossy optical cavity},\ }\href {https://doi.org/10.1080/09500340.2011.599881} {\bibfield  {journal} {\bibinfo  {journal} {Journal of Modern Optics}\ }\textbf {\bibinfo {volume} {58}},\ \bibinfo {pages} {1310} (\bibinfo {year} {2011})}\BibitemShut {NoStop}%
\bibitem [{\citenamefont {Thompson}\ \emph {et~al.}(1992)\citenamefont {Thompson}, \citenamefont {Rempe},\ and\ \citenamefont {Kimble}}]{Thompson1992}%
  \BibitemOpen
  \bibfield  {author} {\bibinfo {author} {\bibfnamefont {R.~J.}\ \bibnamefont {Thompson}}, \bibinfo {author} {\bibfnamefont {G.}~\bibnamefont {Rempe}},\ and\ \bibinfo {author} {\bibfnamefont {H.~J.}\ \bibnamefont {Kimble}},\ }\bibfield  {title} {\bibinfo {title} {Observation of normal-mode splitting for an atom in an optical cavity},\ }\href {https://doi.org/10.1103/PhysRevLett.68.1132} {\bibfield  {journal} {\bibinfo  {journal} {Phys. Rev. Lett.}\ }\textbf {\bibinfo {volume} {68}},\ \bibinfo {pages} {1132} (\bibinfo {year} {1992})}\BibitemShut {NoStop}%
\bibitem [{\citenamefont {J\"ager}\ \emph {et~al.}(2021{\natexlab{a}})\citenamefont {J\"ager}, \citenamefont {Liu}, \citenamefont {Shankar}, \citenamefont {Cooper},\ and\ \citenamefont {Holland}}]{jager2021_1}%
  \BibitemOpen
  \bibfield  {author} {\bibinfo {author} {\bibfnamefont {S.~B.}\ \bibnamefont {J\"ager}}, \bibinfo {author} {\bibfnamefont {H.}~\bibnamefont {Liu}}, \bibinfo {author} {\bibfnamefont {A.}~\bibnamefont {Shankar}}, \bibinfo {author} {\bibfnamefont {J.}~\bibnamefont {Cooper}},\ and\ \bibinfo {author} {\bibfnamefont {M.~J.}\ \bibnamefont {Holland}},\ }\bibfield  {title} {\bibinfo {title} {Regular and bistable steady-state superradiant phases of an atomic beam traversing an optical cavity},\ }\href {https://doi.org/10.1103/PhysRevA.103.013720} {\bibfield  {journal} {\bibinfo  {journal} {Phys. Rev. A}\ }\textbf {\bibinfo {volume} {103}},\ \bibinfo {pages} {013720} (\bibinfo {year} {2021}{\natexlab{a}})}\BibitemShut {NoStop}%
\bibitem [{\citenamefont {J\"ager}\ \emph {et~al.}(2021{\natexlab{b}})\citenamefont {J\"ager}, \citenamefont {Liu}, \citenamefont {Cooper}, \citenamefont {Nicholson},\ and\ \citenamefont {Holland}}]{jager2021_2}%
  \BibitemOpen
  \bibfield  {author} {\bibinfo {author} {\bibfnamefont {S.~B.}\ \bibnamefont {J\"ager}}, \bibinfo {author} {\bibfnamefont {H.}~\bibnamefont {Liu}}, \bibinfo {author} {\bibfnamefont {J.}~\bibnamefont {Cooper}}, \bibinfo {author} {\bibfnamefont {T.~L.}\ \bibnamefont {Nicholson}},\ and\ \bibinfo {author} {\bibfnamefont {M.~J.}\ \bibnamefont {Holland}},\ }\bibfield  {title} {\bibinfo {title} {Superradiant emission of a thermal atomic beam into an optical cavity},\ }\href {https://doi.org/10.1103/PhysRevA.104.033711} {\bibfield  {journal} {\bibinfo  {journal} {Phys. Rev. A}\ }\textbf {\bibinfo {volume} {104}},\ \bibinfo {pages} {033711} (\bibinfo {year} {2021}{\natexlab{b}})}\BibitemShut {NoStop}%
\bibitem [{\citenamefont {J\"ager}\ \emph {et~al.}(2021{\natexlab{c}})\citenamefont {J\"ager}, \citenamefont {Liu}, \citenamefont {Cooper},\ and\ \citenamefont {Holland}}]{jager2021_3}%
  \BibitemOpen
  \bibfield  {author} {\bibinfo {author} {\bibfnamefont {S.~B.}\ \bibnamefont {J\"ager}}, \bibinfo {author} {\bibfnamefont {H.}~\bibnamefont {Liu}}, \bibinfo {author} {\bibfnamefont {J.}~\bibnamefont {Cooper}},\ and\ \bibinfo {author} {\bibfnamefont {M.~J.}\ \bibnamefont {Holland}},\ }\bibfield  {title} {\bibinfo {title} {Collective emission of an atomic beam into an off-resonant cavity mode},\ }\href {https://doi.org/10.1103/PhysRevA.104.053705} {\bibfield  {journal} {\bibinfo  {journal} {Phys. Rev. A}\ }\textbf {\bibinfo {volume} {104}},\ \bibinfo {pages} {053705} (\bibinfo {year} {2021}{\natexlab{c}})}\BibitemShut {NoStop}%
\bibitem [{\citenamefont {Rivero}\ \emph {et~al.}(2023)\citenamefont {Rivero}, \citenamefont {Jr}, \citenamefont {de~França}, \citenamefont {Teixeira}, \citenamefont {Slama},\ and\ \citenamefont {Courteille}}]{Rivero:2023}%
  \BibitemOpen
  \bibfield  {author} {\bibinfo {author} {\bibfnamefont {D.}~\bibnamefont {Rivero}}, \bibinfo {author} {\bibfnamefont {C.~A.~P.}\ \bibnamefont {Jr}}, \bibinfo {author} {\bibfnamefont {G.~H.}\ \bibnamefont {de~França}}, \bibinfo {author} {\bibfnamefont {R.~C.}\ \bibnamefont {Teixeira}}, \bibinfo {author} {\bibfnamefont {S.}~\bibnamefont {Slama}},\ and\ \bibinfo {author} {\bibfnamefont {P.~W.}\ \bibnamefont {Courteille}},\ }\bibfield  {title} {\bibinfo {title} {Quantum resonant optical bistability with a narrow atomic transition: bistability phase diagram in the bad cavity regime},\ }\href {https://doi.org/10.1088/1367-2630/acf954} {\bibfield  {journal} {\bibinfo  {journal} {New Journal of Physics}\ }\textbf {\bibinfo {volume} {25}},\ \bibinfo {pages} {093053} (\bibinfo {year} {2023})}\BibitemShut {NoStop}%
\bibitem [{\citenamefont {Christensen}\ \emph {et~al.}(2015)\citenamefont {Christensen}, \citenamefont {Henriksen}, \citenamefont {Sch\"affer}, \citenamefont {Westergaard}, \citenamefont {Tieri}, \citenamefont {Ye}, \citenamefont {Holland},\ and\ \citenamefont {Thomsen}}]{Christensen:2015}%
  \BibitemOpen
  \bibfield  {author} {\bibinfo {author} {\bibfnamefont {B.~T.~R.}\ \bibnamefont {Christensen}}, \bibinfo {author} {\bibfnamefont {M.~R.}\ \bibnamefont {Henriksen}}, \bibinfo {author} {\bibfnamefont {S.~A.}\ \bibnamefont {Sch\"affer}}, \bibinfo {author} {\bibfnamefont {P.~G.}\ \bibnamefont {Westergaard}}, \bibinfo {author} {\bibfnamefont {D.}~\bibnamefont {Tieri}}, \bibinfo {author} {\bibfnamefont {J.}~\bibnamefont {Ye}}, \bibinfo {author} {\bibfnamefont {M.~J.}\ \bibnamefont {Holland}},\ and\ \bibinfo {author} {\bibfnamefont {J.~W.}\ \bibnamefont {Thomsen}},\ }\bibfield  {title} {\bibinfo {title} {Nonlinear spectroscopy of {Sr} atoms in an optical cavity for laser stabilization},\ }\href {https://doi.org/10.1103/PhysRevA.92.053820} {\bibfield  {journal} {\bibinfo  {journal} {Phys. Rev. A}\ }\textbf {\bibinfo {volume} {92}},\ \bibinfo {pages} {053820} (\bibinfo {year} {2015})}\BibitemShut {NoStop}%
\bibitem [{\citenamefont {Beijerinck}\ and\ \citenamefont {Verster}(1975)}]{beijerinck1975}%
  \BibitemOpen
  \bibfield  {author} {\bibinfo {author} {\bibfnamefont {H.~C.~W.}\ \bibnamefont {Beijerinck}}\ and\ \bibinfo {author} {\bibfnamefont {N.~F.}\ \bibnamefont {Verster}},\ }\bibfield  {title} {\bibinfo {title} {{Velocity distribution and angular distribution of molecular beams from multichannel arrays}},\ }\href {https://doi.org/10.1063/1.321845} {\bibfield  {journal} {\bibinfo  {journal} {Journal of Applied Physics}\ }\textbf {\bibinfo {volume} {46}},\ \bibinfo {pages} {2083} (\bibinfo {year} {1975})}\BibitemShut {NoStop}%
\end{thebibliography}%

\appendix
	\section{Details on the numerical simulation of the mean-field equations\label{App:Numerical}}
In this appendix, we report details on the simulation of Eqs.~\eqref{eq:alpha}-\eqref{eq:meanx}.

We simulate an effective two-dimensional beam that describes the atomic density along the cavity axis ($z$-axis) and the beam axis ($x$-axis). In this description, we ignore optomechanical forces, this assume that the atomic momentum width is much larger than the photon recoil. In addition, we assume that the spatial width of the beam (along the $z$-axis) is much wider than an optical wavelength. This allows us to assume that the beam is spatially homogeneous over many wavelengths and stationary in time.

To simulate the equations of motion we define unitless quantities as follows
\begin{align}
\tilde{x}=&\frac{x}{2w},\\
\tilde{z}=&kz,\\
\tilde{v}_x=&\frac{p_x}{m\bar{v}_x},\\
\tilde{v}_z=&\frac{kp_z\tau}{m},\\
\tilde{t}=&\frac{t}{\tau}.
\end{align}
The equations of motion for the four unitless quantities are
\begin{align}
    \frac{d\tilde{x}}{d\tilde{t}}=\tilde{v}_x,\label{eq:tildex}\\
    \frac{d\tilde{z}}{d\tilde{t}}=\tilde{v}_z.\label{eq:tildez}
\end{align}
The mode-function in Eq.~\eqref{eq:eta} simplifies to
\begin{align}
    \eta(x,z)=\cos(\tilde{z})e^{-4\tilde{x}^2}.
\end{align}
To simulate a flux of atoms, we define the entry position of the atoms at $\tilde{x}=-3/2$. This gives a very weak atom-cavity coupling reduced by $\exp(-9)\approx1.2\times10^{-4}$.
As a first step, we sample the velocities along the $x$ direction. For this we calculate the cumulative distribution to find a particle of velocity $\tilde{v}_x<\tilde{V}_x$ given by
\begin{align}
    F(\tilde{V}_x)=1-e^{-\frac{\pi}{4}\tilde{V}_x^2}.
\end{align}
By sampling a random number, $a\in[0,1]$, we obtain the random velocity
\begin{align}
    \tilde{v}_x=\sqrt{-\frac{4}{\pi}\log(1-a)},
\end{align}
which determines the velocity of the particles that enter the cavity. The average number of particles is then calculated assuming a homogeneous density
\begin{align}
d\bar{n}=N\times\tilde{v}_x\times d\tilde{t}.
\end{align}
In order to sample the actual integer number of particles $dn$ we assume Poisson statistics and sample $dn$ from a Poisson distribution with mean value $d\bar{n}$. After having set the initial values for $\tilde{x}$, $\tilde{v}_x$, and the atom number, we sample the atomic position along the cavity axis $\tilde{z}$ as a random number between $[0,2\pi)$. This relies on the homogeneity of the atomic density over many wavelengths. The velocity $\tilde{v}_z$ is sampled as a Gaussian random variable with mean $\langle\tilde{v}_z\rangle=\delta_0\tau$ and variance $\langle\tilde{v}_z^2\rangle=\delta_D^2\tau^2.$ This completes the sampling of the atoms that enter the cavity in each time step of the simulation. Their dynamics is subsequently simulated by Eqs.~\eqref{eq:alpha}-\eqref{eq:meansjz} and Eqs.~\eqref{eq:tildex} and \eqref{eq:tildez}. In addition, when the trajectories have reached the point $\tilde{x}=3/2$, we remove the trajectory from our simulation. This is the point where we assume that the atoms leave the cavity and have the same weak coupling between atom and cavity that defines the point of entry.

\section{Details on experimental motion along the atomic-beam axis\label{App:B}}
\begin{figure}
        \centering
         \vspace*{.2in}~\\
	\includegraphics[width=\linewidth]{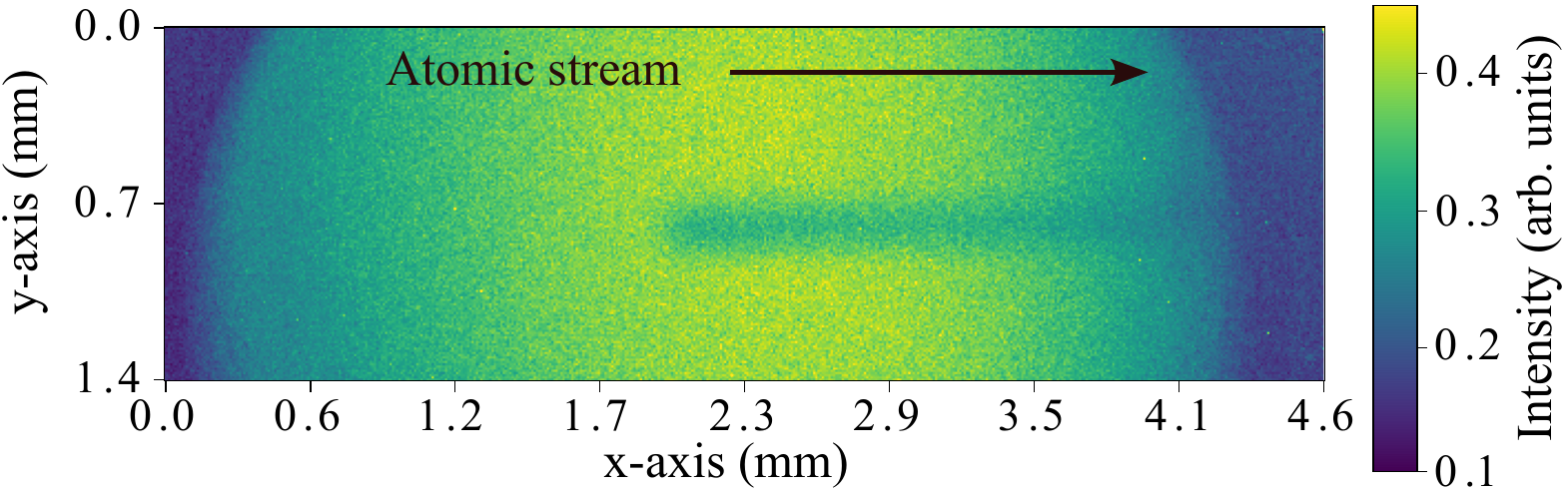}
	\caption{Atomic fluorescence in the cavity seen through a cavity mirror. We couple a~\qty{689}{nm} beam into the cavity. This addresses the atoms at the center of the cavity, roughly~\qty{2}{mm} along the $x$-axis, where they are transferred to $\ket{e}$, transparent to the fluorescence beam. The cut is chosen to highlight the dark stripe due to these flying atoms.   \label{Fig:10}}
\end{figure}

 Here, we outline the experimental procedure for measuring the vertical velocity $v_x$ of the atoms as they transit the cavity mode. To perform this measurement, we exploit the high transparency of our cavity mirrors at the blue transition wavelength, \qty{461}{nm}. We direct a resonant fluorescence beam across the cavity along the $y$-axis. If atoms in $\ket{g} = {}^1S_0$ are present, their fluorescence is collected with a camera, through one of the cavity mirrors. The cavity is resonant with the atomic transition $\ket{g} \rightarrow \ket{e}$, where $\ket{e} = {}^3P_{1, m_j = 0}$. By coupling a resonant~\qty{689}{nm} probe beam into the cavity, the atoms become dark to our blue fluorescence beam, casting a shadow in the fluorescence profile. This is shown in Fig.~\ref{Fig:10}(a) as the appearance of a distinct dark stripe in our fluorescence image.
To measure atomic velocity along the vertical $x$ axis, we modulate both fluorescence and probing beams at a frequency typically within the range of 300 to \qty{400}{kHz}. This frequency is carefully optimized to suit the oven temperature, taking into account the limited optical access provided by the cavity spacer aperture, which is approximately~\qty{4}{mm} wide. 
Upon traversing the cavity mode, an atomic sample is subjected to imaging and probing by the modulated probe beam. Subsequently, it is imaged again after traveling a certain distance determined by the chopping frequency and vertical velocity. By measuring the distance traveled, we can determine $v_x$. 
We use a simple Gaussian fit to estimate the first and second positions and use the width of the second Gaussian as an estimation of the atomic velocity spread. With this method, we measure $v_x = 400\pm\qty{50}{m/s}$ for T =~\qty{773}{K}. This is comparable to the value expected using Eq.~\ref{eq:vxmean}.    

\end{document}